\documentclass[11pt,a4paper]{article}
\pdfoutput=1
\usepackage[pdftex]{graphics}
\usepackage{jheppub}
\usepackage{amsmath,amssymb,amsfonts}
\usepackage{enumitem}
\usepackage{multirow}
\usepackage{array,booktabs}
\usepackage{tikz}
\usepackage{slashed}
\usepackage{verbatim}
\usepackage{float}
\usepackage{graphicx}
\usepackage{subcaption}
\usepackage{empheq}
\usepackage{cancel}
\usepackage{comment}
\usepackage{empheq}
\usepackage{kbordermatrix}

\newcommand{\eq}{\begin{equation}}
\newcommand{\eqe}{\end{equation}}
\newcommand{\eqa}{\begin{eqnarray}}
\newcommand{\eqae}{\end{eqnarray}}

\newcommand{\bn}{\begin{enumerate}}
\newcommand{\en}{\end{enumerate}}
\newcommand{\bl}{\begin{align}}
\newcommand{\el}{\end{align}}
\def\ie{\begin{equation}\begin{aligned}}
\def\fe{\end{aligned}\end{equation}}

\def\jmath{{j}}
\def\bl#1\el{\begin{align} #1 \end{align}}
\def\bg#1\eg{\begin{gather} #1 \end{gather}}

\def\bld#1\eld{\begin{aligned} #1 \end{aligned}}
\def\bgd#1\egd{\begin{gathered} #1 \end{gathered}}

\usetikzlibrary{snakes}
\usetikzlibrary{shapes.misc}
\usetikzlibrary{decorations.markings}

\tikzset{line/.style={line width=0.25mm},
curve/.style={line,smooth,tension=1},
->-/.style={decoration={
  markings,
  mark=at position #1 with {\arrow[>=stealth]{>}}},postaction={decorate}},
-<-/.style={decoration={
  markings,
  mark=at position #1 with {\arrow[>=stealth]{<}}},postaction={decorate}},
}

\title{(Non)-projective bounds on gravitational EFT}
\author[1]{Li-Yuan Chiang}
\author[1,2]{Yu-tin Huang}
\author[3]{Wei Li}
\author[1]{Laurentiu Rodina}
\author[1]{He-Chen Weng} 

\affiliation[1]{Department of Physics and Center for Theoretical Physics, National Taiwan University, Taipei 10617, Taiwan}
\affiliation[2]{Physics Division, National Center for Theoretical Sciences, Taipei 10617, Taiwan}
\affiliation[3]{Department of Physics, Boston University, Boston, MA 02215, USA}
 
\emailAdd{fishbone99999999@google.com, yutinyt@gmail.com, weili17@bu.edu, laurentiu.rodina@gmail.com, albertweng1118@gmail.com}

\abstract{In this paper we study both projective and non-projective constraints on four-dimensional gravitational effective fields theories implied from unitarity, causality and crossing, assuming perturbative UV completions in $M_{\rm pl}$. We derive bounds on the Wilson coefficients of $R^3$ and $D^{2n}R^4$ from its dispersive representation, utilizing both numerical semi-definite programming and analytic geometry analysis. From the former, we derive projective bounds on ratios of couplings and observe accumulation point spectrum populating the boundary of the allowed region. For the latter we consider the non-projective geometry of the EFThedron, which we relate to the known $L$-moment problem in the literature. This allows us to move beyond positivity and incorporate the upper bound from unitarity of the imaginary parts of partial waves. This leads to sharp bounds on individual coefficients, which are of order unity when normalized with respect to the UV scale. Finally, the non-projective geometry also allows us to derive optimal bounds reflecting assumptions of low-spin dominance, improving previous results. We complement the analytic analysis with a simple linear programming approach that validates the bounds.}

\begin{document}
\maketitle

\section{Introduction}
The Einstein-Hilbert action can be viewed as the leading term of the effective description of quantum gravity. Modifications stemming from ultraviolet (UV) completion can in general be parameterized by the presence of higher derivative interactions, suppressed by the cutoff scale of the UV completion. If gravity is UV completed while weakly coupled, then the cutoff scale will be below $M_{\rm pl}$ and one can ask what such an effective theory (EFT) would look like. A canonical example is perturbative string theory, where $g_s=\left(\frac{\ell_{\rm Planck}}{\ell_{\rm s}}\right)^4=\left(\frac{m_{\rm s}}{M_{\rm Planck}}\right)^4\ll 1$. For type-II strings, the tree-level EFT takes the form
\eq
R{-}2\alpha'^{3}\zeta(3)R^4{-}\alpha'^{5}\zeta(5)D^4R^4{+}\frac{2}{3}\alpha'^{6}\zeta^2(3)D^6R^4{-}\frac{1}{2}\alpha'^{7}\zeta(7)D^8R^4{+}\cdots\,.
\eqe 
In the above, we have restrict ourselves to operators that modify the low energy four-graviton amplitude. Thus a sharper version of the question is what are the allowed values for the Wilson coefficients of these operators. Note that without the assumption of weak coupling, the natural scale for the EFT cutoff is simply $M_{\rm pl}$.

It has been pointed out long ago that on general grounds, the coefficient of the leading higher derivative operator is constrained by UV unitarity (and IR causality) to be positive~\cite{Adams:2006sv} (see~\cite{Pham:1985cr, Ananthanarayan:1994hf} for even earlier discussion). More precisely, the leading Wilson coefficient can be written as a dispersive integral of the  imaginary part of the forward amplitude. Since unitarity tells us that the latter is proportional to the total cross-section which is positive, the positivity of the Wilson coefficient follows. Extending the dispersive representation beyond the forward limit and imposing crossing symmetry has led to a steady flow of progress in recent years, uncovering an infinite number of constraints for the infinite series of Wilson coefficients~\cite{Bellazzini:2014waa,Bellazzini:2016xrt, deRham:2017avq, deRham:2017zjm, Bellazzini:2020cot,  Tolley:2020gtv, Caron-Huot:2020cmc, Arkani-Hamed:2020blm, Sinha:2020win,Du:2021byy,Alberte:2021dnj,Zhang:2021eeo}.

Most of the work mentioned above focuses on scalar EFTs. For gravity several new features arise. First, due to the presence of external spins, the expansion basis for the dispersion relation is no longer a unique polynomial. In $D=4$, the three-point amplitude of two massless and one massive spinning state is unique once the helicity and the spin are determined. Gluing two three-point amplitudes for the exchange of a spin-$\ell$ state leads to the Wigner $d$-function~\cite{Arkani-Hamed:2017jhn, Hebbar:2020ukp} (see also~\cite{Arkani-Hamed:2020blm, Bern:2021ppb}), which depends on the spin and the helicity difference in the exchange channel. For $D>4$ since the three-point coupling of two massless gravitons and  an irreducible representation of SO($D{-}1$) is not unique, the expansion is on a matrix basis (see \cite{Caron-Huot:2022jli} for a recent analysis).

Second is the famous ``$t$-pole" problem, referring to the singularity of tree-level graviton exchange $\sim \frac{s^2}{t}$ near the forward limit. There are two consequence, first is that it invalidates the Froissart bound~\cite{Froissart:1961ux}, which limits the forward amplitude to $M(s,0)<s \log^{D-2}s$ at large $s$, allowing a twice subtracted dispersive representation. This can be circumvented by invoking AdS/CFT and viewing the graviton scattering amplitude as the flat-space limit of the bulk S-matrix dual to the four-point function of stress-tensors, one can derive that the amplitude grows as  $<s^2$ in the Regge limit~\cite{Caron-Huot:2017vep,Caron-Huot:2021enk,Chandorkar:2021viw}.  The second is that the dispersive representation is given in terms of an infinite sum over spin partial waves, and a singularity in the forward limit indicates the sum being non-convergence in the forward limit for operators proportional to $s^2$.  This can be addressed by considering  dispersion relations with finite impact parameter~\cite{Caron-Huot:2021rmr}, or assume Regge behaviour at high energies, and hence the contribution above the Regge scale will cancel against the graviton pole leaving behind a finite contribution~\cite{Tokuda:2020mlf}. In this paper we will simply consider operators that are not affected by such convergence issues.

At low energies the gravitational amplitude, for example for $M(1^+, 2^+,3^-,4^-)$, will take the form, 
\eq
M(1^+, 2^+,3^-,4^-)=\langle 12\rangle^4 [34]^4\left(\frac{8\pi G_N}{stu}{+}\alpha_1\frac{1}{s}{+}\alpha_2\frac{tu}{s}{+}\sum_{k,q\geq 0}b_{k,q}s^{k{-}q}t^q\right)
\eqe
where $\alpha_{1,2}$ parameterize the presence of $\phi R^2$ and $R^3$ operators. Note that we've neglected massless loops that would lead to IR divergences and the running of these coefficients. This is valid in our context since we are assuming perturbative UV completions, where massless loop are of higher order in $G_N$ and hence suppressed. The dispersive representation at fixed $t$ with $t\ll 1$, where we've normalized the cut-off scale $M$ to $1$, can be expressed as the following equality:
\eq\label{Dis}
M(1^+, 2^+,3^-,4^-)={-}16\pi[12]^{4}\langle 34\rangle^{4}\;\left(\sum_{a} \frac{d_{0,0}^{\ell_a{=}even}(\theta)}{m_a^{6}} \frac{|p_{a}^{++}|^2}{s{-}m^2_a}{+}\sum_{b} \frac{d_{4,4}^{\ell_b\geq4}(\theta)}{m_a^{6}\cos^{8}(\frac{\theta}{2})} \frac{|p_{b}^{+-}|^2}{{-}s{-}t{-}m^2_b}\right)\,,
\eqe
where the equality is understood by matching both sides under the Taylor series expansion at small $s$ and $t$. The matching is valid for all $b_{k,q}$ since the spinor factors  $\langle 12\rangle^4 [34]^4\sim s^4$ ``shields" the operators from the $t$-pole singularity. Here $p_{a}^{\lambda_1,\lambda_2}$ (where $\lambda_{1,2}=\pm$) corresponds to the coupling constant of the external $(\lambda_1,\lambda_2)$ helicity states to a spin-$\ell_a$ irreducible representation of SO(3). Note that we only require the latter to be an irrep and not a single particle state. Thus it is applicable for both tree and (massive) loop-level UV completions, with the distinction given by whether one has a sum or an integral over the states.

In general, dispersive representations such as eq.(\ref{Dis}) bound the couplings to reside within a convex region. The characterization of the region is can be identified with the Minkowski sum of convex hull of moments~\cite{Bellazzini:2020cot, Arkani-Hamed:2020blm, Chiang:2021ziz, Bellazzini:2021oaj}. Crossing symmetry can be imposed on the EFT couplings in the form of linear constraints, i.e. the couplings at equal $k$ are related and must live on a subplane. Thus the geometric picture is the intersection of such crossing-planes (null planes) with a convex hull, which results in a finite bounded region~\cite{Tolley:2020gtv, Caron-Huot:2020cmc, Arkani-Hamed:2020blm, Sinha:2020win, Bern:2021ppb}. 

The boundaries of such space can be approached either numerically using semi-definite programming (SDP)~\cite{Simmons-Duffin:2015qma, Landry:2019qug}, pioneered for the EFT bootstrap in~\cite{Caron-Huot:2020cmc} and applied to multi-scalar fields in ~\cite{Li:2021lpe} and photon EFTs in~\cite{Henriksson:2021ymi}, or analytically utilizing the known boundaries of the (tensored) moment problem, which is characterize by the total positivity of suitable generalized Hankel matrices. The latter approach was initiated in~\cite{Arkani-Hamed:2020blm} and termed the ``EFThedron", with the detailed boundary structure for external massless scalars investigated in~\cite{Chiang:2021ziz}. The positivity of Hankel matrices in relation to polynomials of multiple zeta values was studied in \cite{Green:2019tpt}. When the ``planar" EFThedron was subject to monodromy relations implied by the presence of a worldsheet CFT, the EFT coefficients was shown to be constrained to the open superstring theory \cite{Huang:2020nqy}. 

In this paper we will explore the constraints on gravitational EFT using both approaches. We begin with the dispersive representation for all helicity sectors of the four-point amplitude, i.e. the all plus, single minus and MHV configuration. Applying SDPB allows us to impose global consistency conditions, which results in bounds on coefficients for $R^3$ and $D^{2n}R^4$. As an example, the ratio of the coefficient for $D^2R^4$ ( $b_{1,0}$) and square of $R^3$ ($b_{1,1}$), both normalized with respect to $R^4$ ($b_{0,0}$), is shown as follows:
$$ \includegraphics[scale=0.6]{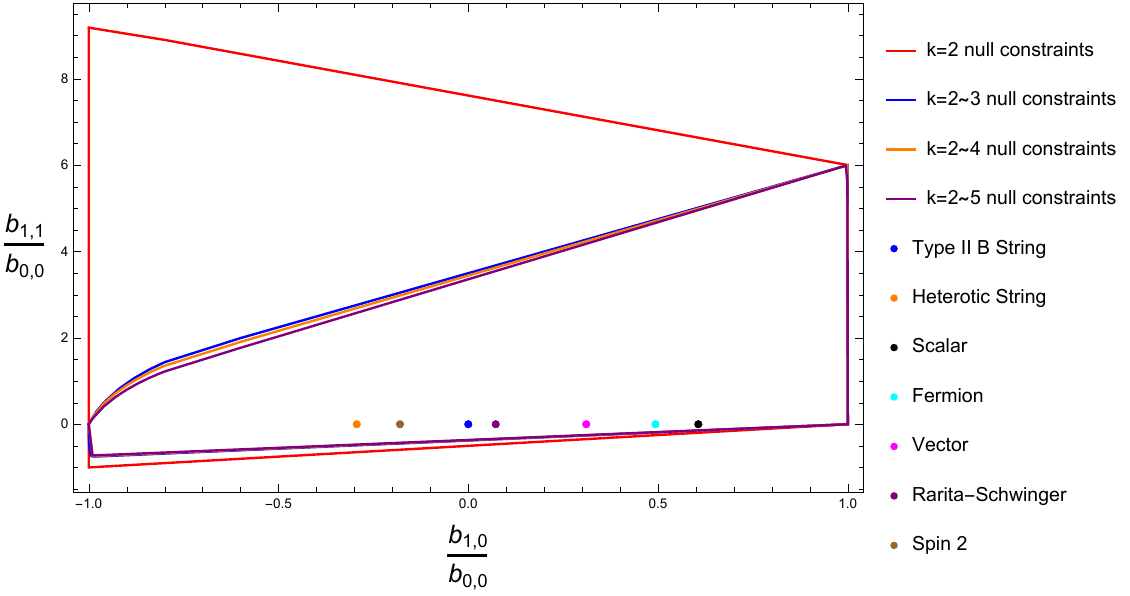}\,.$$
The different colored lines indicate different number of crossing (null) symmetry constraints imposed, where at each derivative order ($k$) there are finite number of them. As one can see, the boundaries converges rapidly already at $k\sim4$. The known theories, including type-II, Heterotic string, as well as low energy effective theory of integrating out massive states, lies on a horizontal line in the plot since for these theories $R^3$ operators are either forbidden by SUSY, or only arises at subleading order in $G_N$.

Global consistency also leads to improved bounds such as for the operators $D^{8}R^4$. There are three distinct on-shell couplings at this order, which can be combined into two ratios  $(\frac{b_{4,2}}{b_{4,0}}, \frac{b_{4,1}}{b_{4,0}})$, we obtain the following plot 
$$\includegraphics[scale=0.6]{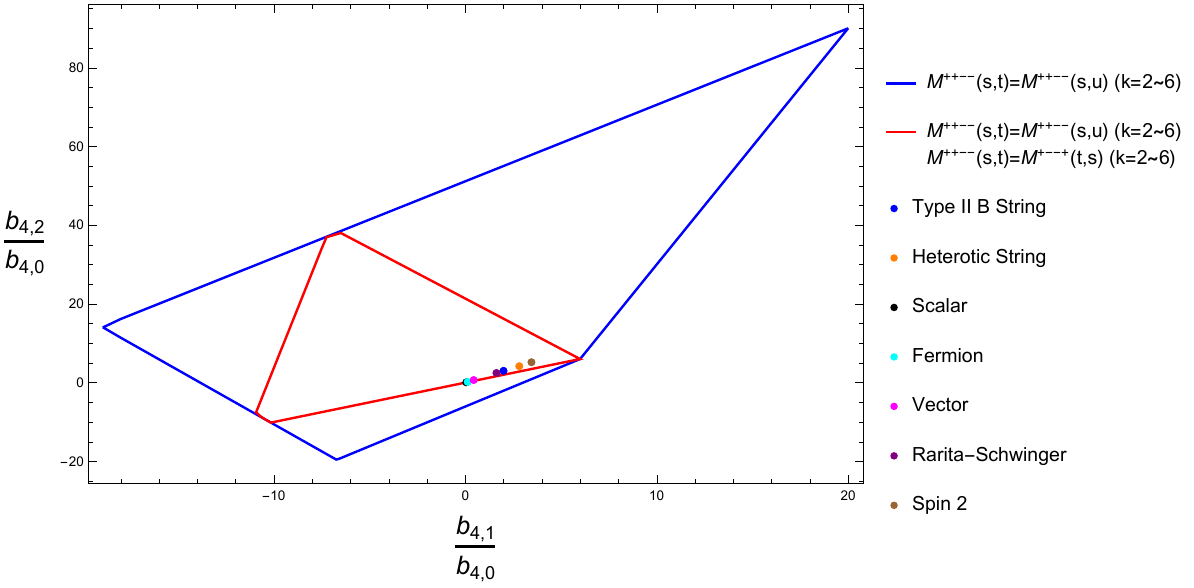}\,.$$
While the space is large compared to small region populated by known theories, it is instructive to view the space in a three-dimensional plot, where the extra coordinate is the ratio of $\frac{b_{4,0}}{b_{0,0}}$. With $\frac{b_{4,0}}{b_{0,0}}$ held fix, the corresponding two dimensional slice takes the form 
$$\includegraphics[scale=0.5]{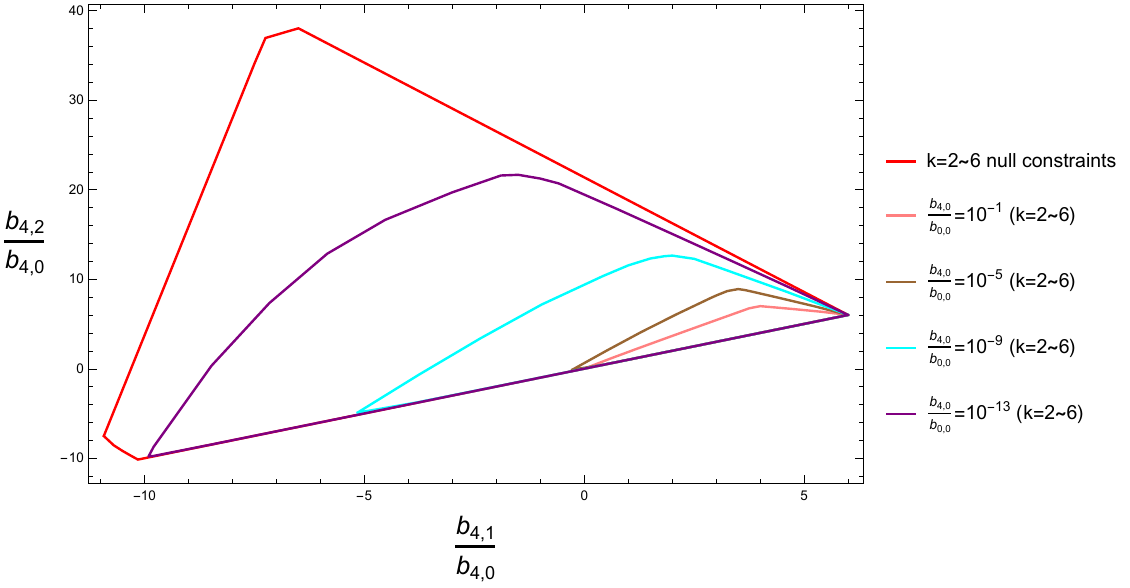}\,.$$
We see that for generic region outside of the known theories, the upper bound is $\frac{b_{4,0}}{b_{0,0}}\ll 1/100$. With the cut-off normalized to $1$, this is dramatic fine-tuning. Thus we see that when viewed in the direction involving couplings of distinct dimensions, the majority of the space is populated by known theories. Indeed this is what we find when consider the 2-d space for $\frac{b_{2,q}}{b_{0,0}}$ in Figure.\ref{b21-b20_projective_space}. 

Using SDPB, from the extremal functional that determines the boundaries, we can read off the spectrum by reading off its zeros. When the boundaries converge as we increase the number of null constraints, special attention is warranted to spectrums which are stable during this process. We find that along the contours, the emergence of ``accumulation point" spectrum, which consists of a tower of higher spin states at the cut-off. This is reminiscent of that found in the scalar EFT bootstrap, where an amplitude that realizes this spectrum was found~\cite{Caron-Huot:2020cmc}.

So far we have only taken into account the positivity of $|p_{a}^{\lambda_1\lambda_2}|^2\equiv \rho^{\lambda_1\lambda_2}_{\ell}(m)$. Unitarity also imposes that it is also bounded from above $0\leq\rho^{\lambda_1\lambda_2}_\ell(m)\leq 2$. The implications of such bounds were explored for scalar EFT in~\cite{Caron-Huot:2020cmc}. Note that the upper bounds are generically considered as non-perturbative unitarity constraint. Here, while we are assuming weak coupling in $M_{\rm pl}$, the UV states can still have self couplings that are non-perturbative, making the exploration of such bounds relevant. Here we will utilize the analytic geometry of the dispersive representation to systematically explore this non-projective constraint. By identifying the space with the Minkowski sum of convex hulls of moments, we can map the constraint $0\leq\rho^{\lambda_1\lambda_2}_\ell(m)\leq 2$ to the ``$L$-moment" problem in mathematical literature~\cite{akhiezer1962some, akhiezer1934fouriersche}, which concerns with finding the sufficient conditions on $a_{k}$ such that it can be represented as
\eq
a_k=\int_{\mathcal{I}} dz\, z^k \rho(z)\,,
\eqe
where $0\leq\rho(z)\leq L$ and $\mathcal{I}$ is some integration domain. Physical couplings can then be understood as related to Minkowski sum of moments with each spin being weighted differently. More precisely, we have
\eq
b_{k,q}=\sum_{i=s, u}\left(\sum_\ell \lambda^{(i)}_{\ell,k,q}\int_0^1 dz\, z^k \rho_\ell(z)\right)\,,
\eqe
where $\lambda^{(s,u)}_{\ell,k,q}$ is a polynomial in spin that depends on $k,q$ and $s$ or $u$-channel for its origin. By utilizing the known boundary solution to the single $L$-moment problem, and the fact that the boundary of a Minkowski sum is contained in the Minkowski sum of boundaries, we can identify the correct boundary within the infinite sum. This non-projective EFThedron is explored in detail in an accompanied paper~\cite{Chiang:2022ltp}, and in this paper we mainly import the results to derive bounds for the gravitational EFThedron. For example, with $k=1$ null constraint $b_{1,1}=0$, we bound $b_{0,0}$ (the coupling for $R^4$) to be 
\eq
0\leq\frac{b_{0,0}}{16\pi^2}\leq \frac{0.337}{M^8}\,,
\eqe   
where $M$ is the UV cutoff. This is consistent with the intuition that the coefficient of higher dimension operators naturally come with $\mathcal{O}(1)$ numbers weighted over the UV cutoff. 

Since crossing symmetry constraints are represented as linear relations amongst Wilson coefficients at fixed $k$, they represent planes in the space of $b_{k,q}$s. Thus to study the implications of multiple null constraints, we need to understand the geometry in high dimension spaces, which becomes tedious at the moment. To describe more general spaces, we introduce a basic linear programming approach, based on discretizing the integral over $z$. In this setup imposing the unitarity bound is trivial, and Mathematica's implementation of FindMaximum/Minimum is sufficiently efficient for our purposes in this paper. We are thus able to verify that the $L$-moment problem geometry indeed gives sufficient conditions where we expect it to. This numerical approach also allows us to incorporate higher order null constraints.

Finally, the map to the $L$-moment problem can also be applied to derive analytical regions for low spin dominance~\cite{Bern:2021ppb} (LSD), which is formulated as the condition 
\eq
 \langle \rho^{\lambda_1\lambda_2}_0 \rangle_k \geq \alpha \langle \rho^{\lambda_1\lambda_2}_\ell \rangle_k,\quad  \forall \ell\neq 0\,,
\eqe
where we defined the mass averaged spectral function,
\eq
\langle \rho^{\lambda_1\lambda_2}_\ell \rangle_k \equiv \int dz\, z^k \rho^{\lambda_1\lambda_2}_\ell(z),
\eqe 
and $\alpha\geq0$ parameterizes the ``degree" of LSD.  We show that this condition can be formulated as an $L$-moment problem, which we can solve exactly for equal $k$ spaces. For example, going back to the projective $k=4$ plot, we carve out the LSD strip as 
$$ \includegraphics[height=2.5in]{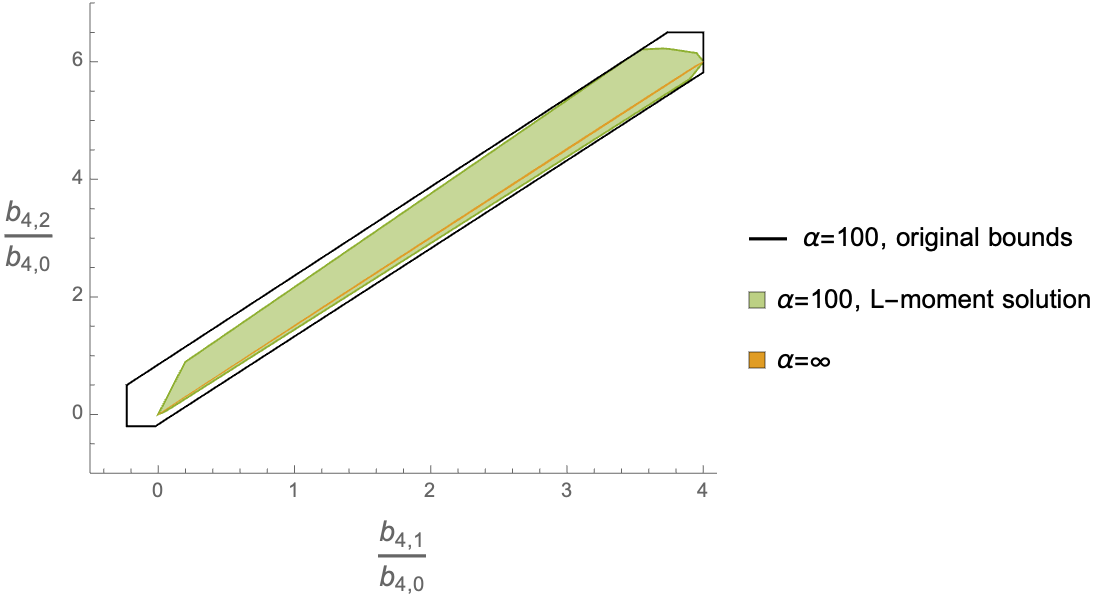}\,.$$  
We see that the optimal boundary derived from the $L$-moment problem for $\alpha=100$ is slightly smaller than that derived in~\cite{Bern:2021ppb}. The linear programming approach is also easily adjusted to incorporate the LSD condition, opening the path to efficiently determining the consequences of this condition on any space, beyond polytopes at equal $k$.

This paper is organized as follows. In Section \ref{Section1} we write the dispersive representation for the various amplitude helicity configurations, as well as for the null constraints originating from crossing or relabeling symmetry. In Section \ref{Section2} we utilize SDP to carve out the projective gravitational S-matrix. We observe that much of the allowed region is inhabited by theories with accumulation points. In Section \ref{Section3} we briefly introduce the non-projective EFThedron and the $L$-moment problem as a way to impose the unitarity bound. We then utilize the constraints to derive non-projective bounds. We also propose a numerical method to impose the unitarity constraint. In Section \ref{Section4} we utilize the non-projective geometry to find the optimal solution to the LSD condition. We similarly compare results to the linear programming approach. We end with Conclusions and outlook in Section \ref{conclusion}.

Note added: during the completion of this manuscript, the authors became aware of the works of~\cite{Caron-Huot:2022ugt}, which contain some overlapping results.

\section{The dispersive representation for helicity amplitudes}\label{Section1}
Let us consider the following low energy effective description of gravity in four-dimensions (following the conventions of ~\cite{Bern:2021ppb}):
\eq\label{EFTL}
S_{EFT}=\int dx^4\;\sqrt{-g}\left( R{+} \frac{\lambda_3}{3!} R^{3}{+}\frac{\lambda_4}{2^3}(R^{2})^2{+} \frac{\tilde\lambda_4}{2^3}(R\tilde{R})^2{+}\cdots\right)\,,
\eqe
where each Riemann tensor and Ricci scalar are normalized with a power of $\frac{2}{\kappa}$, such that it cancels with powers of $\kappa$ in front of the graviton field $h_{\mu\nu}$ when expanding the metric around flat space, $g_{\mu\nu}=\eta_{\mu\nu}{+}\kappa h_{\mu\nu}$. Note that we only list operators that are independent on-shell. The Wilson coefficients $\lambda_n$ are dimensional quantities scaling as $\lambda_n\sim \frac{\#}{M^n}$, where $M$ is the UV cutoff associated with the mass of the UV state. These Wilson coefficients can be matched to the Taylor coefficients of expanding the graviton amplitude in small $s,t$. For a detailed discussion of this matching see ~\cite{Bern:2021ppb, Arkani-Hamed:2020blm}.

At four-points there are three helicity sectors, with all outgoing convention these are the all plus $M^{{+,+,+,+}}$, MHV $M^{{+,+,-,-}}$ and single minus $M^{{+,+,+,-}}$. This implies that the on-shell operators are defined with its power in $s,t$ as well as helicity configuration.\footnote{In relating the Wilson coefficients in the effective action to the Taylor coefficients of the amplitude, there is an additional extra factor of $\kappa^4$ from extracting the four-graviton matrix elements.} Here we will first denote the Taylor coefficients the three sectors, as well as the distinct helicity arrangements within each sector separately. For  example $M^{{+,+,-,-}}$ and $M^{{+,-,-,+}}$ will be denoted as $b,c$ respectively. The purpose for this redundant labelling is that they will have different dispersive representations. The fact that they originate from the same operator, related by simple relabeling, leads to ``crossing symmetry" constraints on the distinct dispersive representations. 

Let us begin with the all plus configuration
\begin{align}\label{EFT1}
M^{{+},{+},{+},{+}}&=\left(\frac{[12][34]}{\langle 12\rangle\langle 34\rangle}\right)^2\left(\sum_{k,q}a_{k,q}s^{k-q}t^q\right),\quad k\geq2,
\end{align}
where there are no pole terms, and the contributions from graviton or dilaton exchange between $R^3$ or $\phi R^2$ only result in polynomials terms. Note that the exponents of $s,t$ are labeled such that $k$ represents the total derivative order ($2k$) of the operator and $q$ is subject to $q\leq k$. Since $q$ counts the power of $t$ it corresponds to the highest spin partial wave associated with the operator. Crossing symmetry relates the distinct coefficients. For the same helicity amplitude, this is the usual permutation invariance 
\begin{equation}\label{++++crossing}
M^{{+},{+},{+},{+}}(s,t)=M^{{+},{+},{+},{+}}(s,{-}s{-}t)=M^{{+},{+},{+},{+}}(t,s),
\end{equation}
which translates to $\sum_{k,q}a_{k,q}s^{k-q}t^q$ being spanned by $\sigma_2=s^2{+}t^{2}{+}u^2$ and $\sigma_3=s^3{+}t^{3}{+}u^3$.

For the MHV configuration, we have:
\begin{align}\label{EFT2}
M^{{+},{+},{-},{-}}&=[12]^{4}\langle 34\rangle^{4}\left(\frac{8\pi G_N}{stu}{+}\alpha_1\frac{1}{s}{+}\alpha_2\frac{t^2}{s}{+}\sum_{k,q}b_{k,q}s^{k-q}t^q\right), \nonumber\\
M^{{+},{-},{-},{+}}&=[14]^{4}\langle 23\rangle^{4}\left(\frac{8\pi G_N}{stu}{+}\alpha_1\frac{1}{t}{+}\alpha_2\frac{s^2}{t}{+}\sum_{k,q}c_{k,q}s^{k-q}t^q\right)\,. \nonumber\\
\end{align}
Here $\alpha_1$ represents possible contributions from $R^2\phi$ where one exchanges a massless scalar, and $\alpha_2$ represents the presence of $R^3$ operators exchanging a graviton. There are two types of crossing identities. First, the exchange symmetry of the same helicity state tells us that 
\begin{align}\label{++--crossing1}
&M^{{+},{+},{-},{-}}(s,t)=M^{{+},{+},{-},{-}}(s,{-}s{-}t)\,.
\end{align}
The above simply implies that $\sum_{k,q}b_{k,q}s^{k-q}t^q$, and its conjugate amplitude denoted as $\sum_{k,q}\tilde{b}_{k,q}s^{k-q}t^q$, are spanned by combinations of $s$ and $(t^2{+}u^2)$. Note that special care is required for imposing the constraint at $k=1$, since it would imply that $b_{1,1}=0$. This  would be true if $R^3$ is absent for example due to supersymmetry, i.e. $\alpha_2=0$. However in the presence of an $R^3$ operator, then it contributes to $M^{{+},{+},{-},{-}}$ as
\eq
-[12]^4\langle 34\rangle^4 \frac{tu}{s}=[12]^4\langle 34\rangle^2 \left(\frac{t^2}{s}+t\right)\,.
\eqe
In this case one has $\alpha_2=b_{1,1}$. Thus in the absence of supersymmetry, the $k=1$ crossing constraint simply implies $b_{1,1}=\alpha_2$, while for supersymmetric theories, one would have an extra $k=1$ constraint $b_{1,1}=0$. Secondly, the fact that different configurations are related by simple relabeling tells us that:
\begin{align}\label{++--crossing2}
&M^{{+},{+},{-},{-}}(s,t)=M^{{+},{-},{-},{+}}(t,s)=M^{{-},{-},{+},{+}}(s,t)=M^{{-},{+},{+},{-}}(t,s)\,,
\end{align}
that is,
\begin{equation}\label{b-c_crossing}
b_{k,q}=c_{k,k{-}q}=\tilde{b}_{k,q}\,.
\end{equation}

For the single minus configuration, we have:
\begin{align}\label{EFT3}
M^{{+},{+},{+},{-}}&=f_{123,4}\left(\frac{\alpha_3}{stu}{+}\sum_{k,q}e_{k,q}s^{k-q}t^q\right), \;&
M^{{+},{+},{-},{+}}&=f_{124,3}\left(\frac{\alpha_3}{stu}{+}\sum_{k,q}f_{k,q}s^{k-q}t^q\right), \nonumber\\
M^{{+},{-},{+},{+}}&=f_{134,2}\left(\frac{\alpha_3}{stu}{+}\sum_{k,q}g_{k,q}s^{k-q}t^q\right), \;&
M^{{-},{+},{+},{+}}&=f_{234,1}\left(\frac{\alpha_3}{stu}{+}\sum_{k,q}h_{k,q}s^{k-q}t^q\right)\,,
\end{align}
where $f_{ijk,l}\equiv([ij][ik]\langle il \rangle)^{4}$. This is in fact permutation invariant with respect to the first three entries. Here there are also two types of crossing symmetry identities. First, we have:
\begin{align}\label{+++-crossing1}
&M^{{+},{+},{+},{-}}(s,t)=M^{{+},{+},{+},{-}}(s,{-}s{-}t)=M^{{+},{+},{+},{-}}(t,s)\,,\nonumber\\
&M^{{+},{+},{-},{+}}(s,t)=M^{{+},{+},{-},{+}}(s,{-}s{-}t)=M^{{+},{+},{-},{+}}(t,s)\,,\nonumber\\
&M^{{-},{+},{+},{+}}(s,t)=M^{{-},{+},{+},{+}}(s,{-}s{-}t)=M^{{-},{+},{+},{+}}(t,s)\,,
\end{align}
this implies that the polynomial in eq.(\ref{EFT3}) is spanned by $\sigma_2$ and $\sigma_3$. Second, we have:
\begin{equation}\label{+++-crossing2}
M^{{+},{+},{+},{-}}(s,t)=M^{{+},{+},{-},{+}}(s,{-}s{-}t)=M^{{-},{+},{+},{+}}({-}s{-}t,t)\,.
\end{equation}

The linear identities amongst the EFT coefficients discussed above are conventionally termed ``null constraints".  For the MHV sector, $b_{k,q}$, previously only the bound arising from the null constraint in eq.(\ref{++--crossing1}) was considered in Ref.~\cite{Bern:2021ppb}.  In this paper we will find that the identification in \label{b-c_crossing} will further improve the bounds on $b_{k,q}$.

The dispersive representation for the EFT coefficients are derived by considering the amplitude at fixed $t$ as a function on the complex $s$ plane. Performing a contour integral around the origin,
\eq\label{Contour}
\frac{1}{2\pi i}\oint \frac{ds}{s^{n{+}1}}M(s,t)\,,
\eqe
picks up the piece of the IR amplitude that is proportional to $s^n$. Here, since the amplitude has a definite spinor pre-factor, we will be mainly interested in the Lorentz invariant dressing function:
\eqa
\frac{M^{{+},{+},{-},{-}}}{8\pi G_N}&=&\left([12]\langle 34\rangle\right)^4\; f(s,t)\nonumber\\
\frac{M^{{+},{+},{+},{-}}}{8\pi G_N}&=&\left([12][13]\langle 14\rangle\right)^4\; h(s, t)\nonumber\\
\frac{M^{{+},{+},{+},{+}}}{8\pi G_N}&=&\left(\frac{[12][34]}{\langle 12\rangle\langle 34\rangle}\right)^2\; g(s, t)\,.\nonumber
\eqae
As the EFT coefficients are the Taylor expansion of $f(s,t)$, $h(s,t)$ and $g(s, t)$, they are identified with the contour integrals of spinor-brackets normalized amplitudes. For example, for the MHV sector, we have
\eq
b_{k, q}=\frac{\partial^q}{\partial t^q}\frac{1}{2\pi i}\oint \frac{ds}{s^{k{-}q{+}1}}\; \frac{M^{{+},{+},{-},{-}}}{ s^4}, \quad  c_{k, q}=\frac{\partial^q}{\partial t^q}\frac{1}{2\pi i}\oint \frac{ds}{s^{k{-}q{+}1}}\; \frac{M^{{+},{-},{-},{+}}}{s^4\sin^8\frac{\theta}{2} }
\eqe
where the extra factors of $s^4$ and $s^4\sin^8\frac{\theta}{2}$ in the denominator are the spinor brackets for $[12]^4\langle 34\rangle^4$ and $[14]^4\langle 23\rangle^4$ respectively.\footnote{Here we use $t=-\frac{s}{2}(1-\cos\theta)$.} Deforming the contour to infinity, the behavior of the amplitude in the Regge limit discussed in the introduction guarantees that the contribution from infinity vanishes for $n\geq2$. This implies that the parameters in the IR amplitude can be identified with the residues or discontinuity on the finite $s$ plane. Now the fact that we are considering $t$ fixed and much smaller than any massive threshold, the only non-analyticity occurs on the real axes, corresponding to physical thresholds. The latter tells us that the contribution corresponds to a product: 
\eq
\textrm{Im}[M(s,t)]|_{s\rightarrow m^2}=\sum_{a} M_3(12 a)M_3^*(34 a)
\eqe 
where $a$ labels the physical states at $s=m^2$, and $M_3$ a three-point amplitude. These can be single or multi-particles states, forming an irrep under the massive little group SU(2) and are thus symmetric spin-$\ell$ states. As discussed in~\cite{Arkani-Hamed:2020blm}, the kinematic dependence of two massless helicity states and one massive spin-$\ell$ interaction is unique, and $M_3(12,a)$ is determined up to a constant which we parameterize as $p^{\lambda_1\lambda_2}_{a}$, with $\lambda_{1,2}$ the helicities of legs $1$ and $2$.

Thus the fixed-$t$ dispersion relations tell us that we can identify~\cite{Bern:2021ppb, Arkani-Hamed:2020blm}:\footnote{Recall that $d^{\ell}_{4,0}(\theta)=d^{\ell}_{0,4}(\theta)=d^{\ell}_{-4,0}(\theta)=d^{\ell}_{0,-4}(\theta)$.}
\begin{align}\label{dispersion++++}
M^{{+},{+},{+},{+}}&={-}\sum_{a}\;p_a^{{+}{+}}p_a^{*--} d^{\ell_a{=}even}_{0,0}(\theta)\left(\frac{m_a^2}{s{-}m_a^2}{+}\frac{m_a^2}{{-}s{-}t{-}m_a^2}\right)\;(n_{kq}\geq2)\nonumber\\
\end{align}
\begin{align}\label{dispersion++--}
 M^{{+},{+},{-},{-}}&={-}s^4\;\left(\sum_{a} \frac{d_{0,0}^{\ell_a{=}even}(\theta)}{m_a^{6}} \frac{|p_{a}^{++}|^2}{s{-}m^2_a}{+}\sum_{b} \frac{d_{4,4}^{\ell_b\geq4}(\theta)}{m_a^{6}\cos^{8}(\frac{\theta}{2})} \frac{|p_{b}^{+-}|^2}{{-}s{-}t{-}m^2_b}\right) \;(n_{kq}\geq0)\nonumber\\
M^{{+},{-},{-},{+}} &= {-}t^4\sum_{a}({-})^{\ell_a} |p_a^{+-}|^2\; \frac{d_{4,{-}4}^{\ell_a\geq4}(\theta)}{m_a^{6}\sin^{8}(\frac{\theta}{2})}\left( \frac{1}{s{-}m_a^2}+\frac{1}{{-}s{-}t{-}m_a^2} \right)\;(n_{kq}\geq2)\nonumber\\
M^{{-},{-},{+},{+}}&={-}s^4\;\left(\sum_{a} \frac{d_{0,0}^{\ell_a{=}even}(\theta)}{m_a^{6}} \frac{|p_{a}^{--}|^2}{s{-}m^2_a}{+}\sum_{b} \frac{d_{4,4}^{\ell_b\geq4}(\theta)}{m_a^{6}\cos^{8}(\frac{\theta}{2})} \frac{|p_{b}^{+-}|^2}{{-}s{-}t{-}m^2_b}\right)\;\;(n_{kq}\geq0)\nonumber\\
\end{align}
\begin{align}\label{dispersion+++-}
M^{{+},{+},{+},{-}}&={-}s^2t^2u^2\sum_a\frac{d^{\ell_a=even{,}\ell_a\geq4}_{0,4}}{\frac{1}{8}m_a^{10}\sin^{4}(\theta)}\; p_a^{++}p_a^{*+-}\left(\frac{1}{s{-}m_a^2}+\frac{1}{{-}s{-}t{-}m_a^2}\right)\;\;(n_{kq}\geq0)\nonumber\\
M^{{+},{+},{-},{+}}&={-}s^2t^2u^2\sum_a\; \left(\frac{d^{\ell_a=even{,}\ell_a\geq4}_{0,-4}}{\frac{1}{4^2}m_a^{10}\sin^{4}(\theta)}\frac{p_a^{++}p_a^{*+-}}{s{-}m_a^2}+\frac{d^{\ell_a=even{,}\ell_a\geq4}_{4,0}}{\frac{1}{4^2}m_a^{10}\sin^{4}(\theta)}\frac{p_a^{+-}p_a^{*--}}{{-}s{-}t{-}m_a^2}\right)\;\;(n_{kq}\geq0)\nonumber\\
M^{{-},{+},{+},{+}}&={-}s^2t^2u^2\sum_a\; p_a^{+-}p_a^{*--} \frac{d^{\ell_a=even{,}\ell_a\geq4}_{-4,0}}{\frac{1}{4^2}m_a^{10}\sin^{4}(\theta)}\left(\frac{1}{s{-}m_a^2}{+}\frac{1}{{-}s{-}t{-}m_a^2}\right)\;\;(n_{kq}\geq0)\nonumber\\
\end{align}
where \(d_{m_1, m_2}^l(\theta)\) is the Wigner-d, and $n_{k,q}\equiv k{-}q$. The equality above is understood in terms of matching coefficients of the Taylor expansion in $s,t$. On the LHS the expansion is matched to the EFT coefficients defined in eq.(\ref{EFT1}),(\ref{EFT2}) and (\ref{EFT3}), with $k,q$ satisfying $n_{k,q}$ greater than the lower bound listed.  This is a reflection that we are working with twice subtracted dispersion rules, i.e. $n\geq2$ in eq.(\ref{Contour}). The RHS represents $s$- and $u$- channel threshold contributions to the fixed-$t$ dispersion relation. Note that we are not making a distinction between tree or loop-level UV contributions, where one is a sum and the other is integral, as the difference does not affect our derivation.

The spectral densities associated with the Wigner-d matrices can be written as a product of couplings between the external gravitons with an internal spin-$\ell$ state. For example, $p_a^{++}$ represents the three point coupling of incoming plus helicity state to an outgoing state with spin $\ell_a$. Note that since our convention is with all momenta incoming, a $({+},{+},{-},{-})$ helicity configuration would correspond to $({+},{+})$ incoming and $({+},{+})$ outgoing gravitons in the $s$-channel,  and the corresponding spectral function is given as $p_a^{++}p_a^{*++}=|p_a^{++}|^2$\,. For later convenience, we give the analytic form of the expansion coefficient for the Wigner-d matrices:
\begin{align}\label{TaylorCoeff}
& \frac{d^{\ell}_{4,4}(\theta)}{\cos^{8}(\frac{\theta}{2})}=\sum_{q}v_{\ell,q}x^q,\quad v_{\ell,q}=\frac{1}{(q!)^2}\prod_{a=1}^q\left[\ell(\ell+1)-(a+4)(a+4-1)\right]\nonumber\\
& \frac{d^{\ell}_{4,-4}(\theta)}{\sin^{8}(\frac{\theta}{2})}=\sum_{q}u_{\ell,q}x^q,\quad u_{\ell,q}=\frac{1}{(q!)(8{+}q)!}\frac{(\ell+4+q)!}{(\ell-4-q)!}\nonumber\\
& \frac{d^{\ell_a}_{4,0}(\theta)}{\sin^{8}(\theta)}=\sum_{q}w_{\ell,q}x^q,\quad w_{\ell,q}=\sqrt{\prod_{a=1}^q\frac{\left[\ell(\ell{+}1){-}(a{+}4)(a{+}4{-}1)\right]}{(4^2 q!(4{+}q)!)^2}\frac{(\ell{+}4{+}q)!}{(\ell{-}4{-}q)!}}\,.
\end{align}
where $\textrm{cos}(\theta)=(1{+}2x)$, and the Taylor expansion in $x$ represents an expansion around the forward limit.

\section{Projective bounds on Wilson coefficients}\label{Section2}
We now combine the constraint of crossing symmetry and the dispersive representation for the EFT couplings. Previous constraint on gravitational EFTs derived from the forward limit dispersion relations, relied on the positivity of the imaginary part of the partial waves~\cite{Bern:2021ppb, Arkani-Hamed:2020blm}, which is not true for helicity configurations where the $t\rightarrow 0$ limit do not correspond to a true ``forward" configuration, i.e. the helicity states for 1,4 cannot be identified in the limit. To harness the constraint from consistency across all configurations, we will consider a semi-definite programming setup. First, the dispersive representation for the EFT couplings can be viewed as a quadratic form in the space of couplings $(p^{++},  p^{--}, p^{+-})$. Since the null constraints are linear in the EFT coefficients, they also take the form of a quadratic form. Thus we schematically have
\begin{equation}\label{sdp_setup}
	 \sum_{a}
	\begin{pmatrix}
		p_a^{++} & p_a^{--} & p_{a}^{+-}
	\end{pmatrix}
	\begin{pmatrix}
		\mathbf{[B_1(\ell_a, m_a) ]_{3\times3}}\\ \mathbf{[B_2(\ell_a, m_a) ]_{3\times3}}\\ \mathbf{[N_1(\ell_a, m_a) ]_{3\times3}} \\  \mathbf{[N_2(\ell_a, m_a) ]_{3\times3}} \\ \vdots 
	\end{pmatrix}
	\begin{pmatrix}
		p_a^{*++} \\ p_a^{*--} \\ p_{a}^{*+-}
	\end{pmatrix}\,= \begin{pmatrix}
		b_1 \\  b_2 \\ 0\\ \vdots 
	\end{pmatrix},
\end{equation}
where $\mathbf{B}_i$s are the ($3\times3$) matrices such that when sandwiched between $\vec{p}_a$s gives the dispersive representation for the EFT couplings $b_i$, while $\mathbf{N}_i$s gives the null constraints. We then consider linear combinations of these quadratic forms to obtain a positive semi-definite matrix for all $\ell_a, m_a$. The combination that optimizes $b_1/b_2$ then provides a maximal/minimal value of this ratio. This optimization procedure is approachable via numerical semi-definite programming (SDP)~\cite{Simmons-Duffin:2015qma, Landry:2019qug}. The application of SDP to EFT bounds was initiated in~\cite{Caron-Huot:2020cmc} for massless scalars, later extended~\cite{Li:2021lpe} to multi-fields and photons~\cite{Henriksson:2021ymi}.

In this section, we will first give a brief review of the setup for SDP, and proceed to analyze the allowed region for EFT couplings in the MHV sector $\{b_{k,q}\}$ as well as all plus sector $\{a_{k,q}\}$. From the dispersive representation in eq.(\ref{dispersion++++}) (\ref{dispersion++--}) and (\ref{dispersion+++-}) one can make two general remarks. First, in constraining the MHV sector, global consistency conditions in the all plus and single minus sector will not impose further constraints. This is because for any spectrum that gives a solution to eq.(\ref{dispersion++--}) satisfying the crossing constraints, one can build a trivial crossing symmetric solution for eq.(\ref{dispersion++++}) and eq.(\ref{dispersion+++-}) by simply requiring that states with non-zero $p_a^{++}$ are orthogonal to those with non-zero $p_{a}^{+-}$ and $p_a^{--}$. This trivially solve the crossing condition since the all plus and single minus amplitude is simply zero. Second, since the all plus and single minus dispersion relation is not in terms of a convex hull, i.e. it is not given by a positive sum of functions. Thus the geometry is trivial. The only non-trivial constraint one can obtain comes from considering their ratios with respect to couplings in the MHV sector. We will observe these features in the explicit analysis.

\subsection{The semi-definite programming setup}
The dispersive representation of the Wilson coefficients in eq.(\ref{dispersion++++}), (\ref{dispersion++--}) and (\ref{dispersion+++-}) are quadratic forms in $p^{++}, p^{--}, p^{+-}$, which can be conveniently organized in the following matrix form:
\begin{equation}\label{b_kq_dispersive_rep}
	\{a_{k,q}, b_{k,q},\cdots \}= \sum_{a}
	\begin{pmatrix}
		p_a^{++} & p_a^{--} & p_{a}^{+-}
	\end{pmatrix}
	\frac{B_{k,q}(\ell_a)}{m_a^{2(k+1)}}
	\begin{pmatrix}
		p_a^{*++} \\ p_a^{*--} \\ p_{a}^{*+-}
	\end{pmatrix}\,,
\end{equation}
where $B_{k,q}(\ell)$ is a $k$-dependent $3\times 3$ matrices whose element is linear in the Taylor coefficients in eq.(\ref{TaylorCoeff}). The null constraint, being linear in the EFT couplings, can be written as a $3\times 3$ matrix form as well:
\begin{equation}\label{n_dispersive_rep}
	n_k(a_{k,q}, b_{k,q},\cdots) = \sum_{a}
	\begin{pmatrix}
		p_a^{++} & p_a^{--} & p_{a}^{+-}
	\end{pmatrix}
	\frac{N_{k}(\ell_a)}{m_a^{2(k+1)}}
	\begin{pmatrix}
		p_a^{*++} \\ p_a^{*--} \\ p_{a}^{*+-}
	\end{pmatrix} = 0\, .
\end{equation}

Now consider bounding  $\frac{M_{gap}^{2(k_2-k_1)}b_{k_2,q_2}}{b_{k_1,q_1}}$, where $M_{gap}$ is the smallest mass of the UV spectrum, which we rescale to $1$ in this paper. Unifying the quadratic forms in eq.(\ref{b_kq_dispersive_rep}) and eq.(\ref{n_dispersive_rep}) into
\begin{equation}\label{Matrix_Vector}
	\vec{F}_{m_a,\ell_a} = 
	\begin{pmatrix}
		\frac{B_{k_1, q_1} (\ell_a) }{ m_a^{2(k_1 + 1)} }\\
		\frac{B_{k_2, q_2} (\ell_a) }{ m_a^{2(k_2 + 1)} }\\
		\frac{N_k(\ell_a)}{m_a^{2(k+1)}}\\
		\vdots
	\end{pmatrix}\,,
\end{equation}
where $B_{k_1, q_1} (\ell_a) $ and $B_{k_2, q_2} (\ell_a)$ are the matrices in eq.(\ref{b_kq_dispersive_rep}) associated with Wilson coefficients $b_{k_1,q_1}$ and $b_{k_2,q_2}$, and ``$\;\;\vdots\;\;$"  represent other null constraints. The constraints to be solved are now written as:
\begin{equation}\label{SDP_form}
    \sum_{\ell_a}
    \begin{pmatrix}
        p_a^{++} & p_a^{--} & p_a^{+-}
    \end{pmatrix}
    \vec{F}_{m_a,\ell_a}
    \begin{pmatrix}
        p_a^{*++}\\
        p_a^{*--}\\
        p_a^{*+-}
    \end{pmatrix}
    =
    \begin{pmatrix}
        b_{k_1,q_1}\\
        b_{k_2,q_2}\\
        0\\
        \vdots
    \end{pmatrix},
\end{equation}
So far $ \vec{F}_{m,\ell}$ is a polynomial in $\ell$ and $1/m$. For latter convenience we will absorb the largest power of $1/m$ into $p_a$, i.e.  
\begin{equation}
    p_a \rightarrow \tilde{p}_a=\frac{p_a}{m_a^{\tilde{k}+1}}, \quad 
    \vec{F}_{m_a,\ell_a} \rightarrow \vec{\tilde{F}}_{m_a,\ell_a} = m_a^{2(\tilde{k}+1)} \vec{F}_{m_a,\ell_a},
\end{equation}
where $\tilde{k} = \textrm{max}\{k_1, k_2, k, ...\}$, keeping in mind that $k_1,k_2$ are the dimensions of the EFT coupling and the remaining $k$s are the null constraints. Writing $m_a^2 = (1+x_a)$, now $\vec{\tilde{F}}_{x_a, l_a}$ consists of polynomials in both $\ell_a$ and $x_a$, with $x_a\geq0$. Finding the upper and lower bounds of $\frac{b_{k_2,q_2}}{b_{k_1,q_1}}$ is a standard semi-definite programming problem. Concretely,
\begin{itemize}
    \item Finding the upper limit of $\left(\frac{b_{k_2,q_2}}{b_{k_1,q_1}}\leq A\right)$ is equivalent to finding a $D{+}2$ vector $\vec{v}$, where $D$ counts the number of null constraints, such that
    \eq
  (0,{-}1,0,\cdots)\cdot\vec{v}=1,\; \&\quad \vec{v}^T\cdot \vec{\tilde{F}}_{x,\ell} \succeq 0\,\quad \forall x \geq 0, \, \ell=0,1,...,\ell_{max}\,.
    \eqe
    Here $\succeq 0$ represents a positive semi-definite ($3\times 3$) matrix, and $\ell_{max}$ represents the spin-truncation, which will require convergence analysis. Once such a $\vec{v}$ is found, from the above one can conclude that:
    \eq
    \vec{v}^T \begin{pmatrix}
        b_{k_1,q_1}\\
        b_{k_2,q_2}\\
        0\\
        \vdots
    \end{pmatrix}= v_1  b_{k_1,q_1} {-}b_{k_2,q_2}\geq0\,,
    \eqe
    where $v_1$ is the first component of $\vec{v}$ and stands for the ratio $\frac{b_{k_2,q_2}}{b_{k_1,q_1}}$. Thus the goal is to search for $\vec{v}$ minimizing $A=v_1$.

        \item Similarly for the lower bound $\left(B\leq\frac{b_{k_2,q_2}}{b_{k_1,q_1}}\right)$ is equivalent to finding $\vec{v}$ such that
       \eq
  (0,1,0,\cdots)\cdot\vec{v}=1,\; \&\quad   \vec{v}^T\cdot \vec{\tilde{F}}\succeq 0\,\quad \forall x \geq 0, \, \ell=0,1,...,\ell_{max}\,.
    \eqe

and one maximizes $B= -v_1$. 
\end{itemize}
Thus the allowed region for $\frac{b_{k_2,q_2}}{b_{k_1,q_1}}$ is the segment $[B,A]$ in $\mathbb{P}^1$. Note that the above discussion is not confined to bounding ratios of EFT coefficients in the same helicity sector. One can similarly bound $\frac{a_{k_2,q_2}}{b_{k_1,q_1}}$, as we will show in the next section. In appendix~\ref{SDPApp} we give an explicit example of bounding couplings  $\frac{b_{4,1}}{b_{4,0}}$, as well as generalizing to bounding multiple ratios. 

While SDP can give us the optimal bound on projective spaces, it can also give us information to the allowed spectrum on the boundary. After one have found the optimal vector $\vec{v}_{opt}$ in the SDP problem, one can dot the vector back in to eq.(\ref{SDP_form})
\begin{equation}\label{spectrum_explain}
 \sum_{\ell_a}
    \begin{pmatrix}
        p_a^{++} & p_a^{--} & p_a^{+-}
    \end{pmatrix}
    \left(\vec{v}_{opt}^T\vec{F}_{m_a,\ell_a}\right)
    \begin{pmatrix}
        p_a^{*++}\\
        p_a^{*--}\\
        p_a^{*+-}
    \end{pmatrix}
    =\vec{v}_{opt}^T
    \begin{pmatrix}
        b_{k_1,q_1}\\
        b_{k_2,q_2}\\
        0\\
        \vdots
    \end{pmatrix}\geq0.
\end{equation}
$\left(\vec{v}_{opt}^T\vec{F}_{m_a,\ell_a}\right)$ is a linear combination of $\vec{F}_{m_a,\ell_a}$ and it is a semi-definite positive three-by-three matrix, therefore each term in the summation must be greater or equal to zero. At the extremal point the right-hand side is zero. Thus in order for a set of $p_a$s to be non-zero, it must coincide with the zeros of $\text{det}\left(\vec{v}_{opt}^T\vec{F}_{m_a,\ell_a}\right)$. Therefore just by finding the double roots of $\text{det}\left(\vec{v}_{opt}^T\vec{F}_{m_a,\ell_a}\right)$ will give us the spectrum that is allowed on the boundary.\footnote{In principle, a viable solution can have any of the zeros populated, not necessarily all of them. But since the absence of a state is no different than an extremely small $p_a$, we will not make such distinction.}

\subsection{Explicit Region Analysis}\label{explict_projective_region}

\noindent\textbf{The $\left({\scriptstyle \frac{b_{1,0}}{b_{0,0}}, \frac{b_{1,1}}{b_{0,0}}}\right)$ space }

\begin{figure}[H]
    \centering
    \includegraphics[scale=0.6]{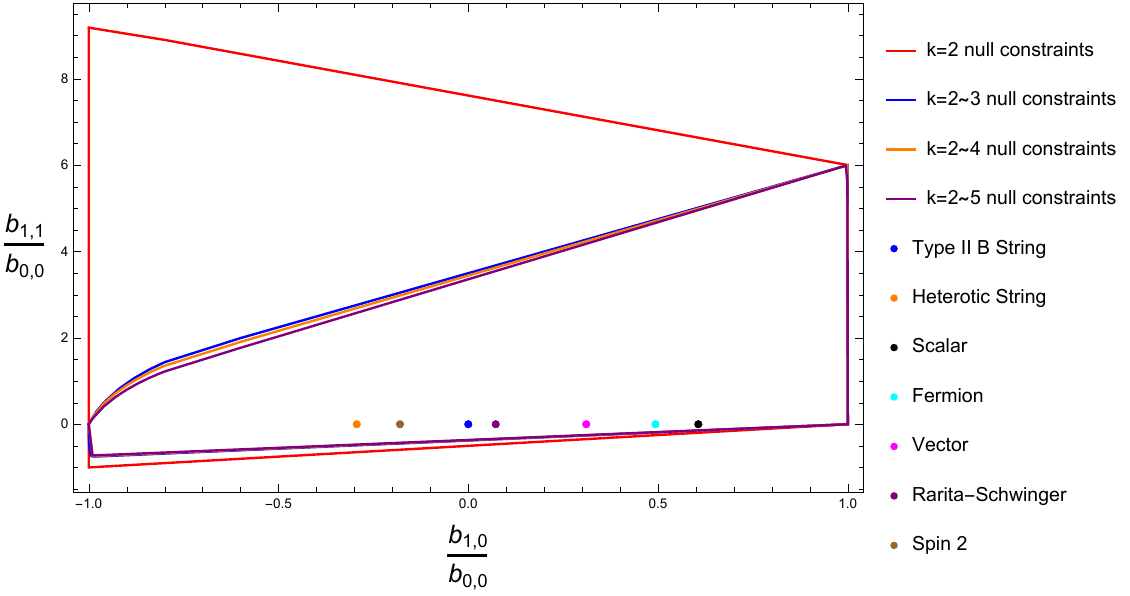}
    \caption{The massive loop amplitudes and string amplitudes lies on the line $b_{1,1}=0$. Imposing $k=2\sim5$ null constraints, we obtained a two-sided bound: $-0.72399\leq\frac{b_{1,1}}{b_{0,0}}\leq5.99643$. }
    \label{b11-b10_projective_space}
\end{figure}

Let us begin by bounding the square of the coefficient of $R^3$, as well the coefficient of $D^2R^4$, normalized by that of $R^4$. 
For the $\left({\scriptstyle \frac{b_{1,0}}{b_{0,0}}, \frac{b_{1,1}}{b_{0,0}}}\right)$ space, the result is shown in Figure \ref{b11-b10_projective_space}. First note that for $\frac{b_{1,0}}{b_{0,0}}$, there are no non-trivial bounds beyond that from power counting, i.e. \(\left| M_{\text{gap}}^2 \frac{b_{1,0}}{b_{0,0}} \right| \leq 1\). More precisely, from eq.(\ref{dispersion++--}) it can be seen that 
\begin{equation}
b_{k,0} = \sum_a \frac{|p_{a}^{++}|^2 + (-1)^k |p_{a}^{+-}|^2}{m_a^{2(k+1)}}, 
\end{equation}
which means that
\begin{align} 
    \left| M_{\text{gap}}^2 b_{1, 0} \right| = \left| \sum_{a} \frac{|p_a^{++}|^2 - |p_a^{+-}|^2}{m_a^2} \left( \frac{M_{\text{\text{gap}}}^2}{m_a^2} \right) \right| \leq \sum_a \frac{|p_a^{++}|^2 + |p_a^{+-}|^2}{m_a^4} = b_{0,0}.
\end{align}
For the square of the coefficient of $R^3$, $b_{1,1}$, we obtain a two-sided bound on its ratio with respect to $R^4$, 
\eq
-0.72399\leq\frac{b_{1,1}}{b_{0,0}}\leq5.99643\,.
\eqe
We see that the known theories lie on the line $b_{1,1}=0$,  due to either the $R^3$ operator forbidden by supersymmetry, or that it is of subleading order in $M_{\rm pl}$. 

\noindent\textbf{The $\left({\scriptstyle \frac{b_{4,1}}{b_{4,0}}, \frac{b_{4,2}}{b_{4,0}}}\right)$ space }

\begin{figure}[H]
   \begin{center}
    \includegraphics[scale=0.6]{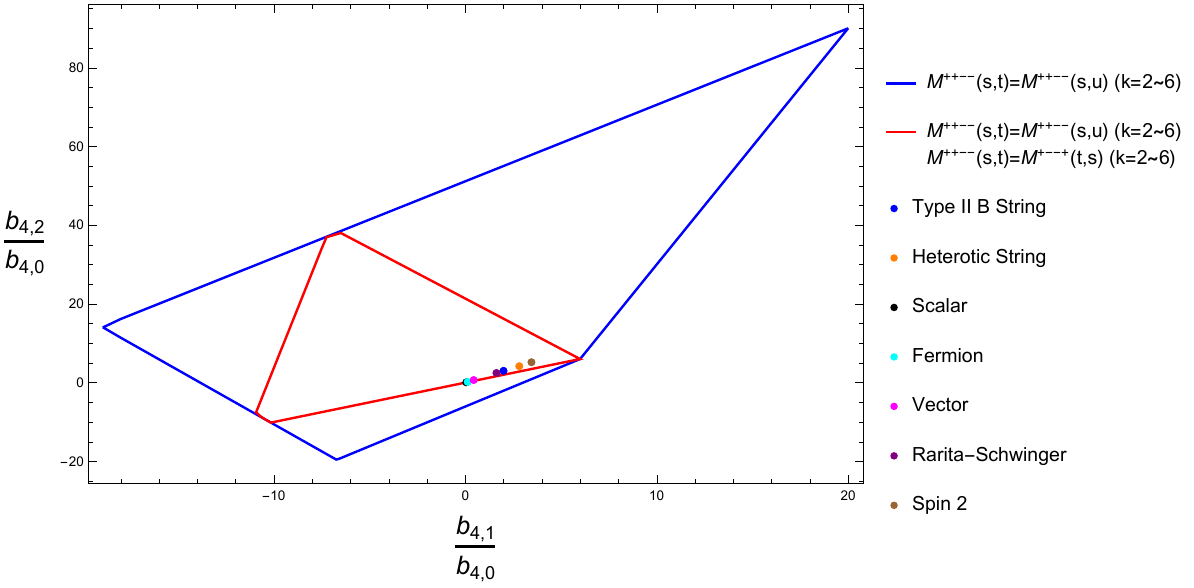}
    \caption{Projective space of \(b_{4,2}\) vs. \(b_{4,1}\), normalized by \(b_{4,0}\). The blue and red lines represent different boundaries of the allowed region when different sets of crossing conditions indicated above are considered. Known theories are all located close to a corner.}
    \label{b42-b41_projective_space}
    \end{center}
\end{figure}

Let us now consider the space for ratios of couplings of the same dimensions, i.e. with $k=4$ for operators $D^8R^4$. The null constraints from  eq.(\ref{++--crossing1}) up to $k=6$ are given as:
\begin{align}\label{tunullconstraintk=4plot}
&k{=}1:\, b_{1,1}{=}0\quad (\text{when } R^3\text{ is absent}), \quad k{=}2:\, b_{2,1}{-}b_{2,2}{=}0,\nonumber\\
&k{=}3:\, b_{3,1}{-}b_{3,2}{=}b_{3,3}{=}0, \quad k{=}4:\, b_{4,3}-2b_{4,4}{=}b_{4,1}{-}b_{4,2}{+}2b_{4,4}{=}0,\nonumber\\
&k{=}5:\, b_{5,3}{-}2b_{5,4}{=}b_{5,1}{-}b_{5,2}{+}b_{5,1}{=}b_{5,5}{=}0,\nonumber\\
&k{=}6:\, b_{6,5}{-}3b_{6,6}{=}b_{6,3}{-}2b_{6,4}{+}5b_{6,6}{=}b_{6,1}{-}b_{6,2}{+}b_{6,4}{-}3b_{6,6}{=}0\,.
\end{align}
Once again the null constraint for $k=1$ is present only when there is no $R^3$ coupling. When $R^3$ is present than its coupling is simply related to $b_{1,1}$. We also have the following constraints from eq.(\ref{++--crossing2}):
\begin{align}\label{stnullconstraintk=4plot}
b_{k,q}{=}c_{k,k-q}\quad \forall\, 2\leq k,\, 0\leq q \leq k-2.
\end{align}
In Figure \ref{b42-b41_projective_space}, the blue contour represents the original result in~\cite{Bern:2021ppb}, where only the null constraints in eq.(\ref{++--crossing1}) at $k=4$ were considered. We find that the space is unchanged when including up to $k=6$ null constraints. However, if we include the null constraint at $k=4$ in eq.(\ref{++--crossing2}) the space drastically reduces to the red contour. The inclusion of null constraints in other helicity sectors does not further reduce the space, as expected from the discussion at the beginning of this section. The known theories, plotted above, are located near a single corner.\footnote{Here for known theories, we are including the EFT coefficients obtained by integrating out minimally coupled massive spin-$s$ fields, with $s\leq2$, at one loop computed in~\cite{Bern:2021ppb}. While these are at the same order as the pure graviton loops that we are suppressing, they certainly give a unitary dispersive representation and hence must be inside our bounds.}

\begin{figure}[H]
    \centering
    \includegraphics[scale=0.35]{bk4_k4k0ratio.pdf}  \includegraphics[scale=0.35]{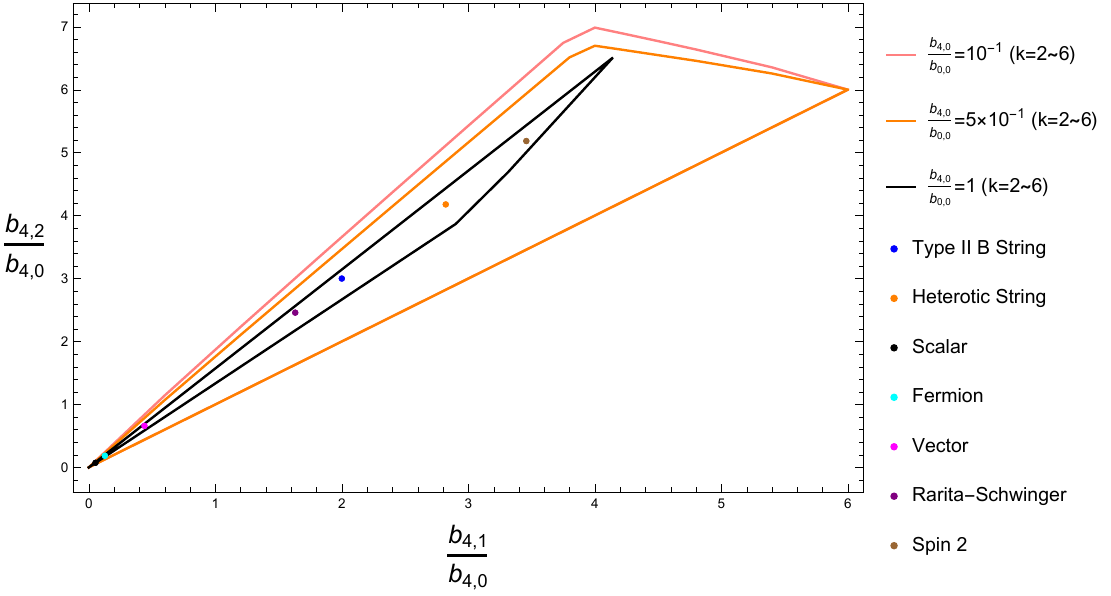}
    \includegraphics[scale=0.35]{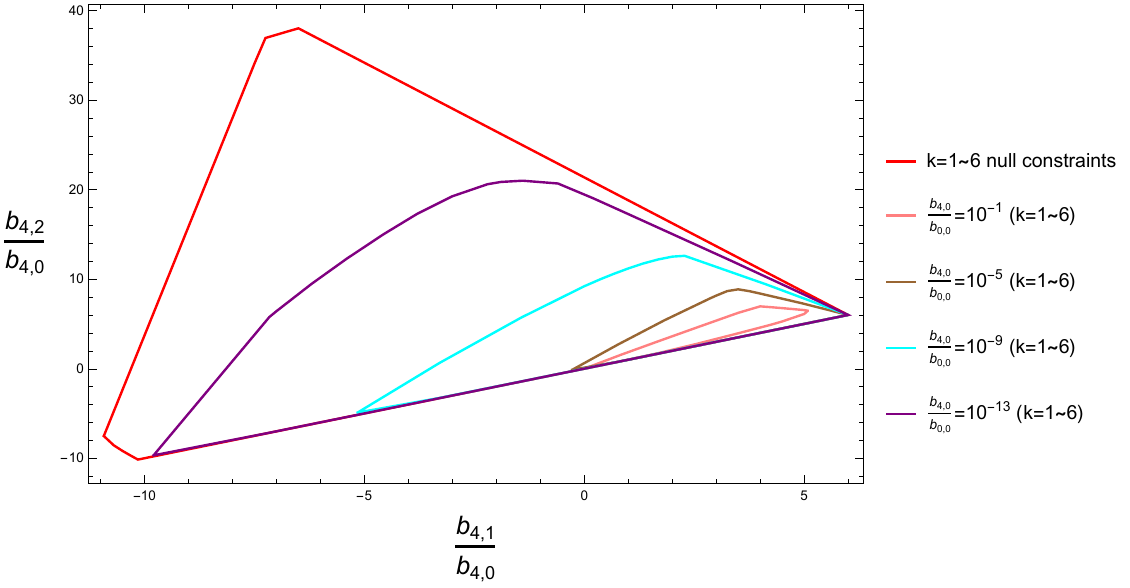}  \includegraphics[scale=0.35]{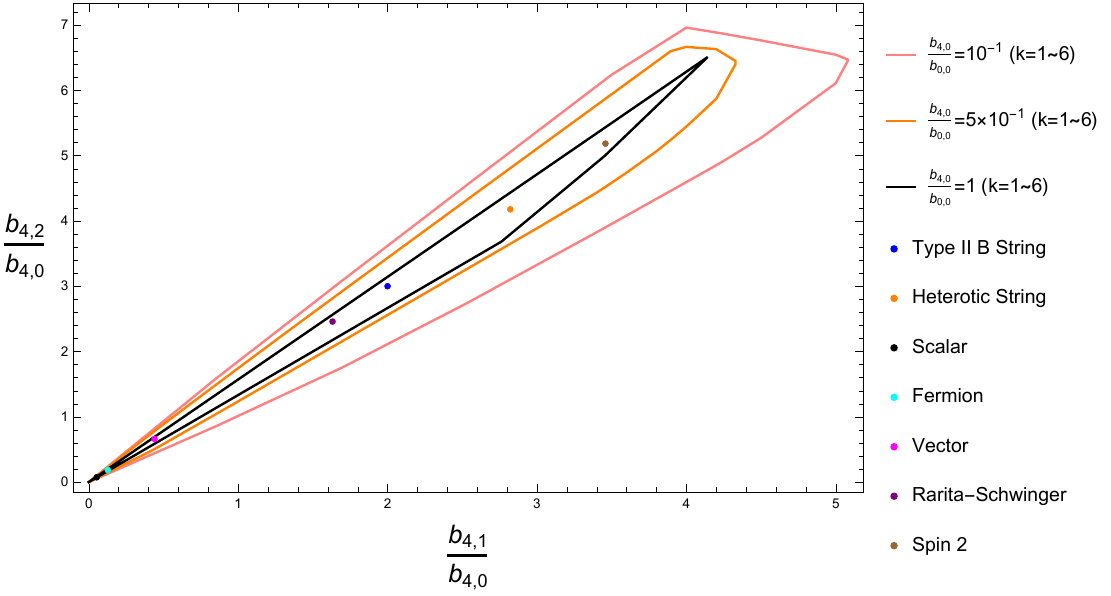}
    \caption{The LHS represents the projective space of \(b_{4,2}\) vs. \(b_{4,1}\), normalized by \(b_{4,0}\). The red line is the same red boundary in Figure \ref{b42-b41_projective_space}, while the boundary enclosed by the red boundary is the projective boundary fixed at various different $\frac{b_{4,0}}{b_{0,0}}$ ratios. On the RHS zooms in on the space for $\frac{1}{10}\leq\frac{b_{4,0}}{b_{0,0}}$, which shows all physical theories lies within. } 
    \label{b40-b41-b42_ratio}
\end{figure}

Note that while the allowed region is relatively large compared to the known theories, when viewed from higher dimensional space where extra couplings are considered, most of the region turns out to be relatively thin, i.e. in the 3D region is ``tortilla" shaped. We consider the contour for extremal $\left(\frac{b_{4,1}}{b_{4,0}},\frac{b_{4,2}}{b_{4,0}}\right)$ with $\frac{b_{4,0}}{b_{0,0}}$ held fixed. Since we are setting the cut-off to be $1$, the natural value for $\frac{b_{4,0}}{b_{0,0}}$ from naive dimensional analysis is $\mathcal{O}(1)$. The result is shown in Figure \ref{b40-b41-b42_ratio} and it is easily seen that much of the region is corresponds to $\frac{b_{4,0}}{b_{0,0}}$ being extremely fine-tuned ($\leq 10^{-5}$). If we require that $\frac{1}{10}\leq\frac{b_{4,0}}{b_{0,0}}$ then the allowed region already shrinks to the lower right corner. In fact the physical theories populate the region where $1=\frac{b_{4,0}}{b_{0,0}}$.\footnote{Due to the mass gap set to 1, $\frac{b_{4,0}}{b_{0,0}}\leq1$. }

\begin{figure}[h]
    \centering
    \begin{subfigure}[h]{0.45\textwidth}
        \includegraphics[width=\textwidth]{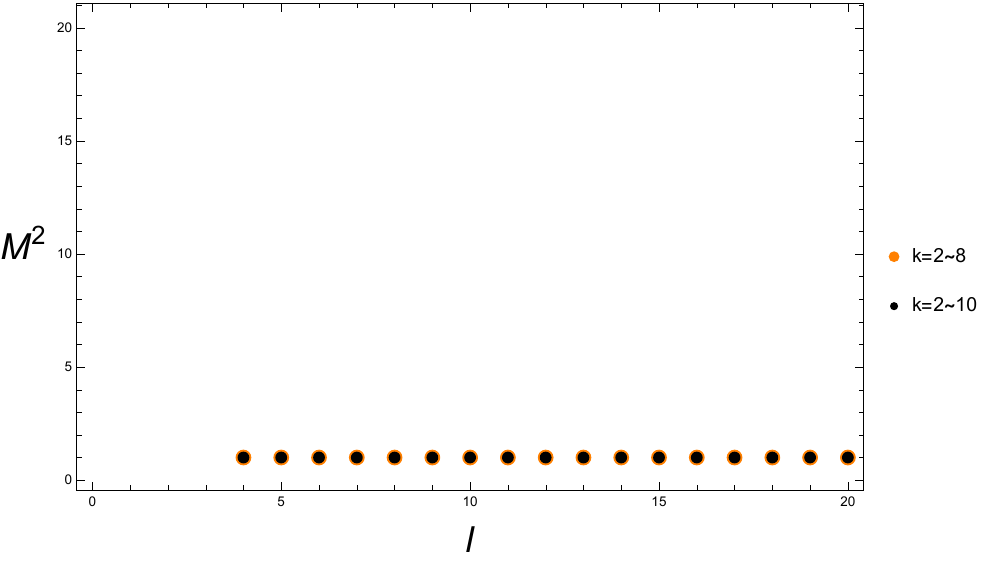}
        \caption{$\frac{b_{4,0}}{b_{0,0}}=1$}
        \label{ratiotenminus1}
    \end{subfigure}
    \begin{subfigure}[h]{0.45\textwidth}
        \includegraphics[width=\textwidth]{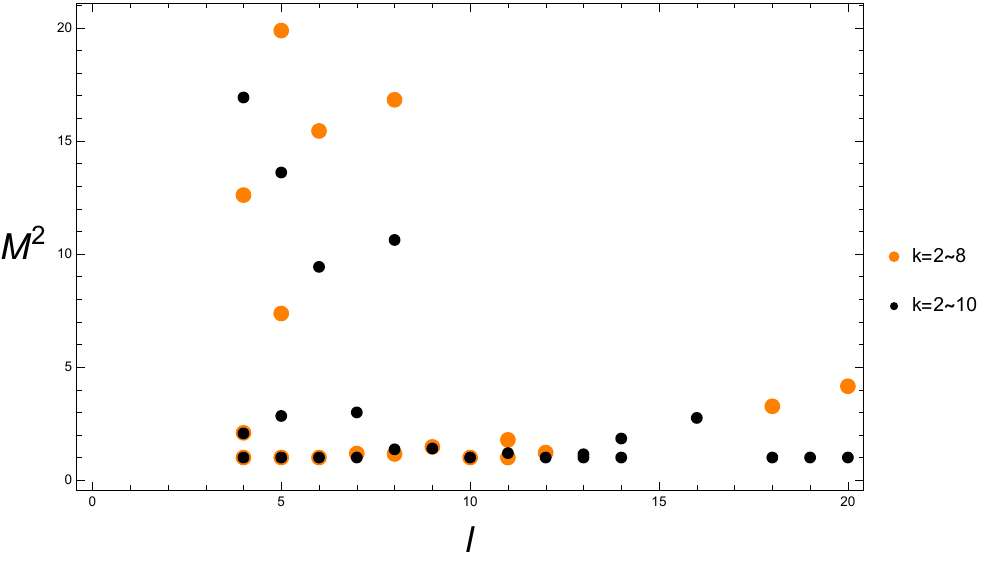}
        \caption{$\frac{b_{4,0}}{b_{0,0}}=10^{-1}$}
        \label{ratiotenminus1}
    \end{subfigure}
    \begin{subfigure}[h]{0.45\textwidth}
        \includegraphics[width=\textwidth]{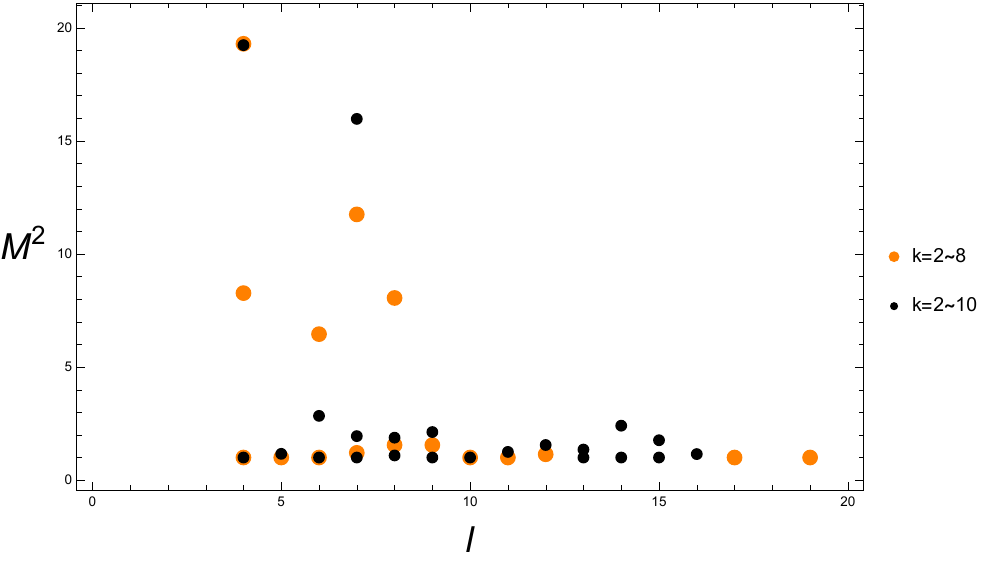}
        \caption{$\frac{b_{4,0}}{b_{0,0}}=10^{-5}$}
        \label{ratiotenminus5}
    \end{subfigure}
    \begin{subfigure}[h]{0.45\textwidth}
        \includegraphics[width=\textwidth]{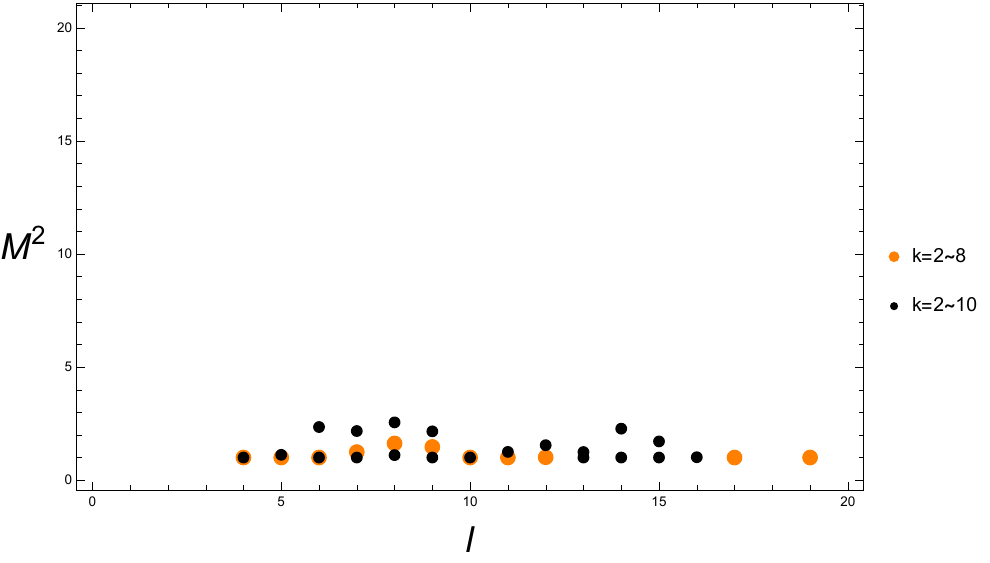}
        \caption{$\frac{b_{4,0}}{b_{0,0}}=10^{-9}$}
        \label{ratiotenminus9}
    \end{subfigure}
    \begin{subfigure}[h]{0.45\textwidth}
        \includegraphics[width=\textwidth]{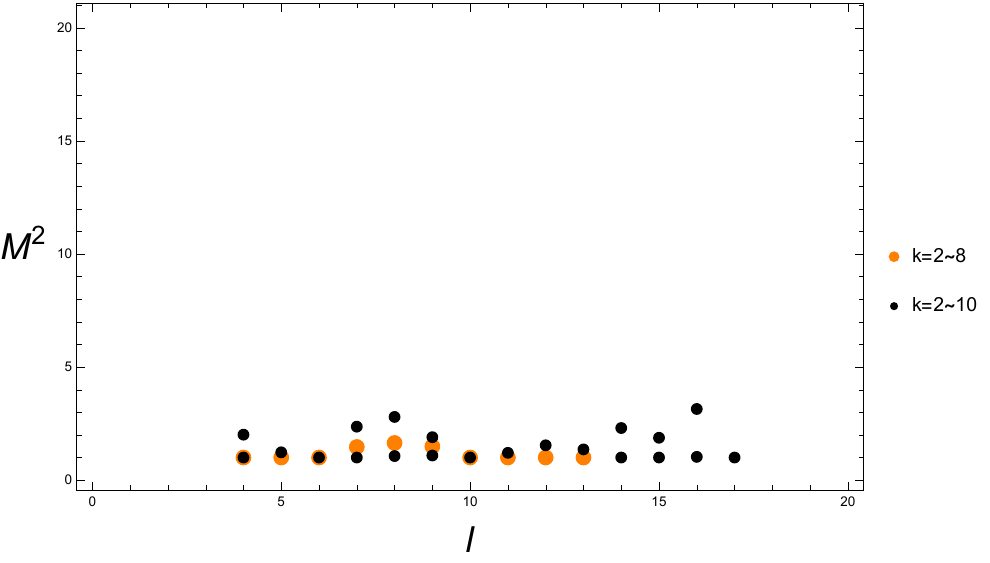}
        \caption{$\frac{b_{4,0}}{b_{0,0}}=10^{-13}$}
        \label{ratiotenminus13}
    \end{subfigure}

    \caption{Boundary spectra that correspond to various $b_{4,0}/b_{0,0}$ ratios with $\frac{b_{4,1}}{b_{4,0}}=1$ held fixed. For each ratio, we compare the zeros under $k=2,\cdots, 8$ null constraints with that under $k=2,\cdots, 10$. We see that in general the zeros trend towards the gap value $M=1$ for each spin. The convergence is faster for larger values of $b_{4,0}/b_{0,0}$ compared to the fine-tuned ones. }
    \label{fig:my_label}
\end{figure}

As mentioned previously, SDPB allows us to infer the boundary spectrum by studying the double-zeros of the optimal functional. However, as one increases the number of null constraints, the zeros will in general shift. Thus it is interesting to study the zeros that are stable under the increase of null constraint. With fixed $\frac{b_{4,1}}{b_{4,0}}=1$, we examine the u-channel spectrum for each $\frac{b_{4,0}}{b_{0,0}}$ value, shown in fig.\ref{fig:my_label}. We show the spectrum with null constraints $k=2\sim 8$ compared to that with $k=2\sim 10$ imposed. We see multiple states with different spin but the same mass $M^2=1$ are `stable". Comparing fine-tuned values of $\frac{b_{4,0}}{b_{0,0}}$ and those closer to $1$, we see that there are more double zeros that tend to being stable at $M^2=1$.

\noindent\textbf{The $\left({\scriptstyle \frac{b_{2,0}}{b_{0,0}}, \frac{b_{2,1}}{b_{0,0}}}\right)$ space }

\begin{figure}[H]
    \centering
    \includegraphics[scale=0.6]{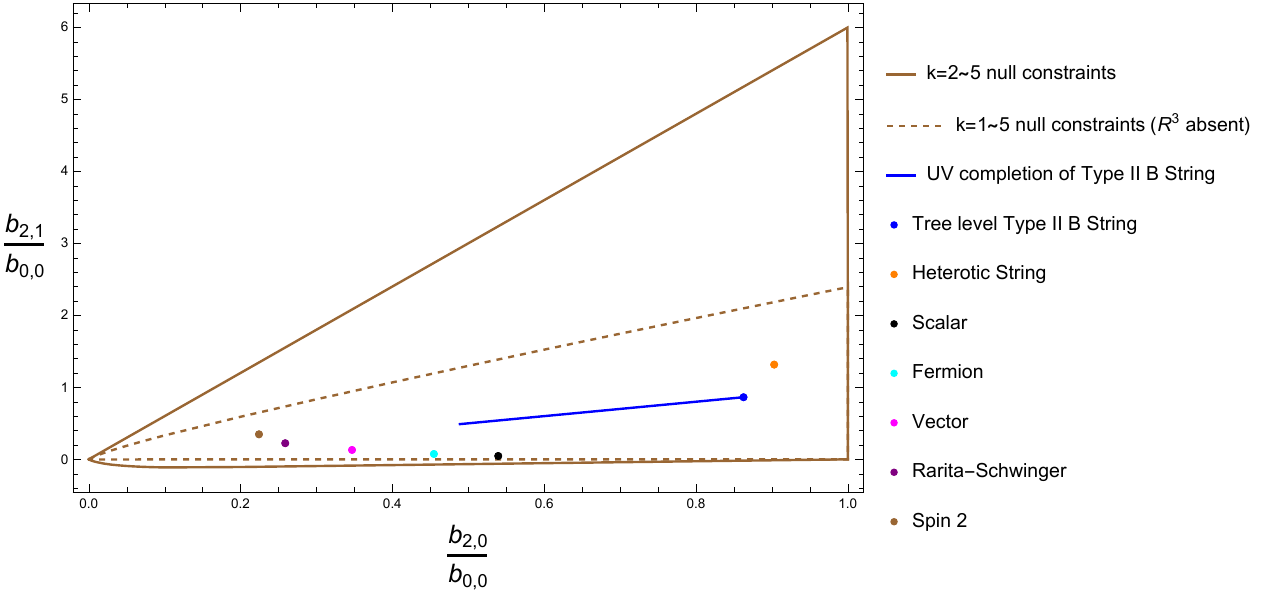}
    \caption{Projective space of \(b_{2,1}\) vs. \(b_{2,0}\), normalized by \(b_{0,0}\). The dashed line represent the case that $R^3$ operator is absent and the solid line represent the contrary. Note that the two string theories are far away from other theories, unlike the equal $k$ space in Figure \ref{b42-b41_projective_space}.}
    \label{b21-b20_projective_space}
\end{figure}

Let us now consider $D^4R^4$ operators normalized by $R^4$ coefficients, where the difference in dimension is made up by the mass gap $M_{gap}$ which is set to 1 in the plots. The result is shown in Figure \ref{b21-b20_projective_space}, where up to $k=5$ null constraints are imposed and they are listed in eq.(\ref{tunullconstraintk=4plot}) and eq.(\ref{stnullconstraintk=4plot}). We use solid lines to delineate the region allowing an $R^3$ operator, and dashed lines the $R^3$ operator is absent.

The string theory values and field theory completions occupy different corners of the region and spans much of the space, unlike the $\left(\frac{b_{4,1}}{b_{4,0}},\frac{b_{4,2}}{b_{4,0}}\right)$ space which are occupied in a tight corner of the space. 
The blue line in the plot is given by the non-perturbative UV completions in type-II B, where the coefficients of $R^4$, $D^4R^4$ and $D^6R^4$ are fixed by S-duality~\cite{Green:1999pu, Green:2014yxa}. These operators are perturbatively non-renormalized, and thus loop corrections are absent. In particular, the $k=0$ and $k=2$ Wilson coefficients are given by the non-holomorphic Eisenstein series, $(b_{0,0}, b_{2,1}){=}(\frac{2\zeta(3)E_{{\scriptstyle \frac{3}{2}}}(\tau)}{\text{Im}[\tau]^{3/2}},\frac{2\zeta(5)E_{{\scriptstyle \frac{5}{2}}}(\tau)}{\text{Im}[\tau]^{5/2}})$, where $\tau$ is the complex moduli, and $b_{2,0}{=}b_{2,1}$ due to supersymmetry. The two ends are given by weak coupling point and $\tau=-0.5+1.328i$.

Note that the minimum of $\frac{b_{2,1}}{b_{0,0}}$ appears to be approaching 0 when $k{=}1$ null constraint, $b_{1,1}=0$, is imposed, indicating an emergent positive bound for the coefficient $b_{2,1}$. We listed below the minimal value of $\frac{b_{2,1}}{b_{0,0}}$ with respect to different null constraints imposed and combine them with the extremal spectrum at spin truncation $\ell_{max}{=}100$, while other extremal values are also listed together:
\begin{center}
\begin{tabular}{|c|c|c|}
\hline
Null constraints & min $\frac{b_{2,1}}{b_{0,0}}$ & Extremal spectrum at min $\frac{b_{2,1}}{b_{0,0}}$: $\{\ell,\, (m/M_{\text{gap}})^2\}$\\
\hline
$k=1\sim 2$ & -0.0979 & s-channel: No spectrum.\\ && u-channel: \{4, 4.4938\}, \{5, 1\}, \{6, 1\}, \{100, 1\}\\
\hline
$k=1\sim 3$ & -0.0085 & s-channel: No spectrum. u-channel: \{4, 8.9331\},\\ && \{5, 3.2489\}, \{6, 1\}, \{7, 1\}, \{99, 1\}, \{100, 1734.5777\}\\
\hline
$k=1\sim 5$ & -0.0017 & s-channel: \{6, 1\}, \{8, 1.5448\}. u-channel: \{4, 1\}, \{4, 25.7889\}, \{5, 1\},\\ &&  \{5, 4.7799\}, \{6, 1\}, \{6, 1.5317\}, \{7, 1.1212\}, \{8, 1\}, \{10, 1.1452\}\\
\hline
Null constraints & max $\frac{b_{2,1}}{b_{0,0}}$ & Extremal spectrum at max $\frac{b_{2,1}}{b_{0,0}}$: $\{\ell,\, (m/M_{\text{gap}})^2\}$\\
\hline
$k=1\sim 5$ & 2.3783 & s-channel: \{2, 1\}, \{4, 1.3004\}, \{6, 1\}, \{6, 1.9816\}, \{8, 1.1354\}.\\ && u-channel: \{4, 1\}, \{6, 1\}, \{7, 1.1983\}, \{8, 1.7313\}, \{9, 1\},\\ && \{10, 1\}, \{99, 1\}, \{99, 392.1765\}.\\
\hline
Null constraints & max $\frac{b_{2,0}}{b_{0,0}}$ & Extremal spectrum at max $\frac{b_{2,0}}{b_{0,0}}$: $\{\ell,\, (m/M_{\text{gap}})^2\}$\\
\hline
$k=1\sim5$ &1& s-channel: \{$\ell,1$\}, $0\leq\ell\leq100,\, \ell \in\text{even}$. u-channel: \{$\ell,1$\}, $4\leq\ell\leq100$.\\
\hline
\end{tabular}
\end{center}
For the minimal value of $\frac{b_{2,0}}{b_{0,0}}=0$, there is no spectrum. Finally, notice that the extremal spectrums also allow an accumulation of state on the mass gap, which is similar to what is seen at the $\left(\frac{b_{4,1}}{b_{4,0}},\frac{b_{4,2}}{b_{4,0}}\right)$ space.

\noindent\textbf{The $\left({\scriptstyle \frac{a_{4,0}}{b_{0,0}}, \frac{a_{5,1}}{b_{0,0}}}\right)$ space }

\begin{figure}[H]
    \centering
    \includegraphics[scale=0.6]{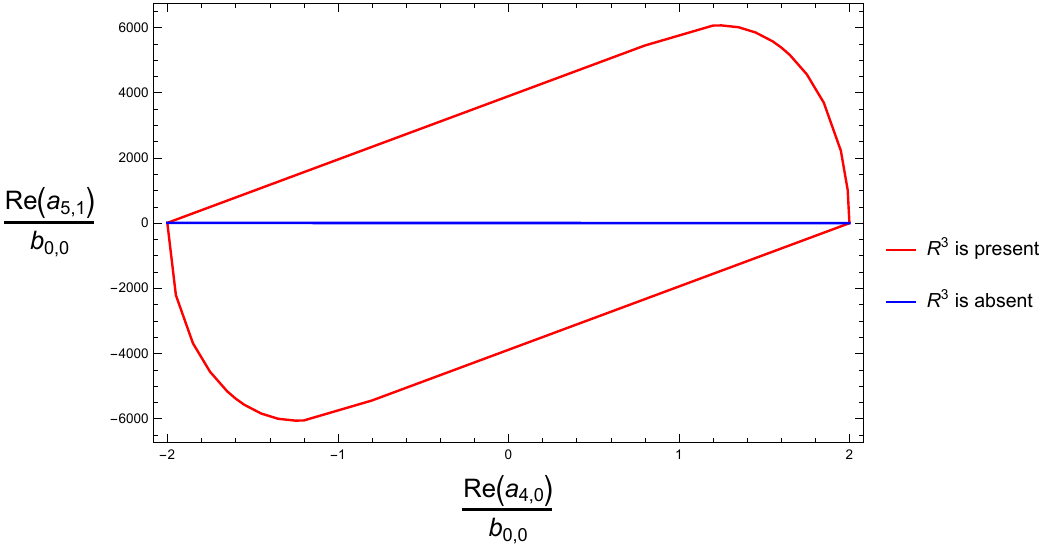}
    \caption{Projective space of \(b_{5,1}\) vs. \(b_{4,0}\), normalized by \(b_{0,0}\). There is a dramatic difference between the red and blue boundary, and the only difference in null constraints is $b_{1,1}=0$ which corresponds to the absence of $R^3$ operator.}
    \label{a51-a40_projective_space}
\end{figure}

\begin{figure}[H]
    \centering
    \includegraphics[scale=0.6]{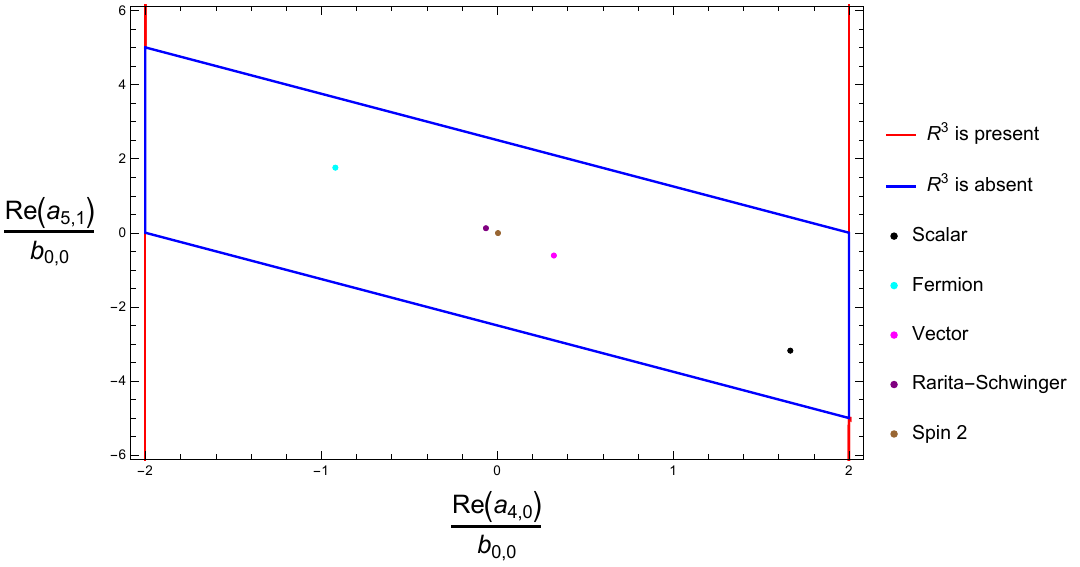}
    \caption{The zoom-in plot of Figure \ref{a51-a40_projective_space}. The field theory completions sit inside the allowed region}
    \label{a51-a40_projective_space_zoom}
\end{figure}

As discussed in the beginning of this section, while the dispersion relations for the all-plus sector do not yield a convex hull, the global constraint allows us to bound the ratio of $a_{k,q}$ with respect to $b_{k,q}$, where the latter enjoys a positive geometry. From eq.(\ref{dispersion++++}), one observes that the coefficients \(a_{k,q}\) are in general complex. We are only able to bound the real part of \(a_{k,q}\), for the matrices in eq.(\ref{sdp_setup}) should be symmetric. Here we consider bounding $\left(\frac{\Re(a_{4,0})}{b_{0,0}},\frac{\Re(a_{5,1})}{b_{0,0}}\right)$. Note that the ratio $\frac{a_{4,0}}{b_{0,0}}$ is independent of the cutoff since they are of the same mass-dimension. The result is given in Figure \ref{a51-a40_projective_space}. The relevant null constraints for $a_{k,q}$ from eq.(\ref{++++crossing}) are given as:
\begin{equation}
k{=}4:\, -2a_{4,0}+a_{4,1}{=}-3a_{4,0}+a_{4,2}{=}0\quad k{=}5:\, a_{5,0}{=}-2a_{5,1}+a_{5,2}{=}-2a_{5,1}+a_{5,3}{=}0.
\end{equation}
There is a huge difference between the red and blue contour, where the only difference with respect to null constraints is that the blue contour assumes the absence of $R^3$ operator (additional null constraints: $b_{1,1}=0$). A zoom-in figure is shown in Figure \ref{a51-a40_projective_space_zoom}, where only the field theory completions are plotted, as supersymmetry forces such amplitude to be absent for type-II and heterotic string. The field theory completion lies inside the allowed region and far from the boundary.

\section{Non-projective bounds on Wilson coefficients}\label{Section3}
So far the bounds we have derived are projective in nature, i.e. in terms of ratios of coupling. This is because in the dispersive representation for the EFT couplings we have only imposed that the imaginary part of the amplitude satisfy consistent factorization, 
\eq\label{41}
{\rm Im}[M]\sim\sum_{\ell} {\rm Im}[a_{\ell}] P_{\ell},\quad \rho_\ell (s)\equiv {\rm Im}[a_{\ell}]=\sum_a p^{\lambda_{in}}_{\ell_a} p_{\ell_a}^{*\lambda_{out}}\,.
\eqe
As the null constraints are always linear in $\rho_\ell (s)$, any global rescaling of $\rho_\ell (s)\rightarrow \lambda \rho_\ell (s)$ will yield a solution, and thus only the relative size of the couplings are meaningful.

In the configuration where the zero scattering angle limit ($t=0$ in the $s$-channel) corresponds to a true forward scattering, i.e. $a, b\rightarrow a, b$, eq.(\ref{41}) implies a positive imaginary part. However,  $p_a$ cannot be arbitrarily large. To see that it is bounded from above consider the scattering amplitude expanded on some orthogonal basis:
\eq
M(s,t)=16\pi \sum_{\ell\;{\rm even}}\;\;(2\ell{+}1) a_{\ell}(s) P_{\ell}(1{+}{\scriptstyle \frac{2t}{s}}) \,,
\eqe
where $ P_{\ell}(1{+}{\scriptstyle \frac{2t}{s}})$s are normalized such that
\eq
\int_{-1}^1 dx P_{\ell}(x) P_{\ell'}(x)=\frac{2\delta_{\ell,\ell'}}{2\ell{+}1}\,.
\eqe
Note that both $d^\ell_{0,0}(\theta)$ and $d^{\ell}_{4,4}(\theta)$ satisfies the above normalization. Unitarity then implies (we follow the convention \cite{Correia:2020xtr})
\eq
2{\rm Im}[a_{\ell}]\geq |a_{\ell}|^2 \;\rightarrow 0\leq {\rm Im}[a_{\ell}]=\sum_a|p_a^\lambda|^2\leq 2\,.
\eqe
This upper bound removes the projective degree of freedom. In~\cite{Caron-Huot:2020cmc} such constraints were incorporated to derive bounds for the coefficient of $(\partial \phi)^4$ yielding 
\eq
0\leq\frac{g_2}{(4\pi)^2}\leq \frac{0.794}{M^4}\,.
\eqe
In a companion paper ~\cite{Chiang:2022ltp}, we will introduce a systematic way of imposing the upper bound for $\rho_\ell (s)\equiv{\rm Im}[a_{\ell}]$ for scalar EFTs. Here we will apply a similar analysis for the gravitational EFT. In the previous numerical analysis, we have seen that the consistency with other helicity structure do not constrain the MHV couplings. Thus it suffices to study the geometry imposed by eq.(\ref{EFT2}) and eq.(\ref{dispersion++--}). To simplify the analysis, we will focus on the dispersion relation for $b_{k,q}$ couplings.  

Starting with the integral dispersive representation for the low energy amplitude valid for $k\geq0$, 
\begin{align}
M^{{+},{+},{-},{-}}(s,t)&=[12]^4\langle34\rangle^4\left(\sum_{k,q}\; b_{k,q}s^{k{-}q}t^q\right)\nonumber\\
&={-}16\pi [12]^4\langle34\rangle^4\sum_{\ell} (2\ell{+}1)\int_{M^2}^{\infty} \frac{dm^2}{\pi m^2} \frac{1}{m^8}\left(\frac{\rho_{\ell}^{{++}}(m^2)d^{\ell}_{0,0}(\theta)}{s/m^2{-}1}{+}\frac{\rho_{\ell}^{{+-}}(m^2)d^{\ell}_{4,4}(\theta)}{\cos^{8}(\frac{\theta}{2})(u/m^2{-}1)}\right)\,,
\end{align}
where $\rho^{{++}}_{\ell}(s)$ and $\rho^{{+-}}_{\ell}(s)$ represent ${\rm Im}[a_{\ell}]$ for $s$ and $u$-channel discontinuities respectively. Note that here we have restored the sum in its integral form to keep track the powers of $m$. Identifying the coefficients of the Taylor expansion on both sides, we have the dispersive representation of $b_{k,q}$.
\eqa\label{bkqdef}
b_{k,q}&=&\int^\infty_{M} \frac{d m^2}{m^2} \frac{1}{(m^{2})^{k{+}4}}\left(\sum_{\ell\in even}\tilde{\rho}_{\ell}^{{++}}(m^2)\lambda^{(s)}_{\ell,q}{+}\sum_{\ell'\geq4}\tilde{\rho}_{\ell'}^{{+-}} (m^2)\lambda^{(u)}_{\ell',k,q}\right)\nonumber\\
&=&\frac{1}{(M^{2})^{k{+}4}} \int^1_{0} \frac{d z}{z} z^{k{+}4}\left(\sum_{\ell\in even}\tilde{\rho}_{\ell}^{{++}}(M^2/z)\lambda^{(s)}_{\ell,q}{+}\sum_{\ell'\geq4}\tilde{\rho}_{\ell'}^{{+-}}(M^2/z)\lambda^{(u)}_{\ell',k,q}\right)\,,
\eqae
where $0\leq \tilde{\rho}^{++}\equiv \leq 32$, $0\leq \tilde{\rho}^{+-}\equiv \leq 16$, and we've introduced $z=M^2/m^2$. Setting the cut-off to be $1$, $0\leq z\leq1$. Here $\lambda^{(s, u)}_{\ell,k,q}$ are polynomials in $\ell,k,q$. 

We can view the geometry of $b_{k,q}$ as the Minkowski sum of the two geometries arising from each channel, 
\eq\label{sudeco}
b_{k,q}=b^{(s)}_{k,q}+b^{(u)}_{k,q}\,.
\eqe
Each geometry itself can in turn be thought of as a Minkowski sum of independent single $L$-moments, rescaled by the factors $\lambda_{\ell,k,q}$.
\eq
b_{k,q}^{(s,u)}=\sum_{\ell} \lambda^{(s,u)}_{\ell,k,q} b_k
\eqe
where the single $L$-moments
\eq
b_k=\int_0^1\rho(z) z^{k+3}dz, \quad 0\le \rho\le L
\eqe
are well studied and have known boundaries. We will provide a simple derivation of these boundaries in the next section.

Finally, the crossing constraints are sub-planes in the space of couplings $b_{k,q}$, as they reflect linear relations. Their intersection with the moment geometry will give bounds on the couplings.

As performing Minkowski sums weighs heavily in our construction, we will begin with the simple example of performing Minkowski sums of vectors, which is already the relevant geometry for the space of couplings with equal $k$. The resulting space is a convex polytope. The main theme is determining its boundary, where one utilizes the following fact: \textit{ the boundary of a Minkowski sum is contained in the Minkowski sum of the individual boundaries}. Thus the task simply boils down to determining which of the latter is a true boundary. After equal $k$, we will generalize to the space of couplings involving different $k$s, where the boundaries involved are curved, and can be obtained from the convex polytope by taking a continuous limit. We will verify the validity of the analysis by numerical methods involving linear programming.

\subsection{The non-projective EFThedron}
In this subsection, we will study the non-projective geometry of the gravitational EFT for the MHV sector. 
\subsubsection{Minkowski sum of vectors, and the equal $k$ space}
Let us consider the simplest example of Minkowski sum of spaces: the sum of $d$-dimensional vectors
\eq\label{PolytopeA}
{\bf b}=\sum^N_{i=1} \rho_i {\bf v}_i\,\quad 0\leq \rho_{i}\leq 1 \,.
\eqe
We will further assume that the vectors are ordered in the sense that 
\eq
\langle {\bf v}_{i_1}{\bf v}_{i_2}\cdots {\bf v}_{i_d}\rangle\geq0,\quad \forall i_1<i_2<\cdots<i_d\,,
\eqe
where $\langle \cdots \rangle$ represents the determinant of the $d\times d$ matrix formed by the ordered vectors. In 2-dimensions, this amounts to the statement that the vectors span a half-plane:
$$\includegraphics[scale=0.4]{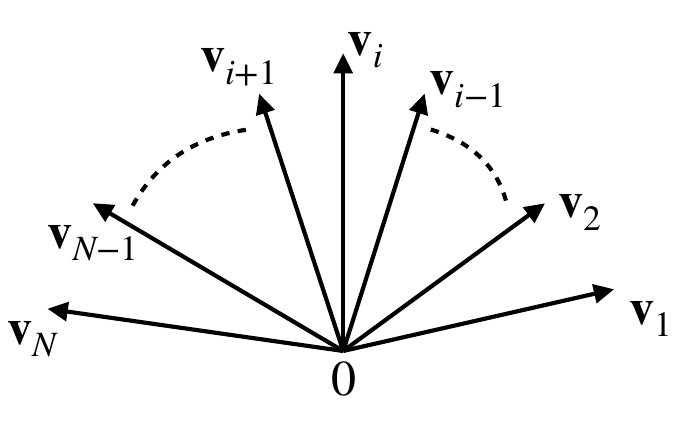}\,.$$
Now we consider the space defined by eq.(\ref{PolytopeA}). It is easy to see that the boundary is simply given by sums of consecutive vectors. For example for $N=3$ we have 
$$\includegraphics[scale=0.45]{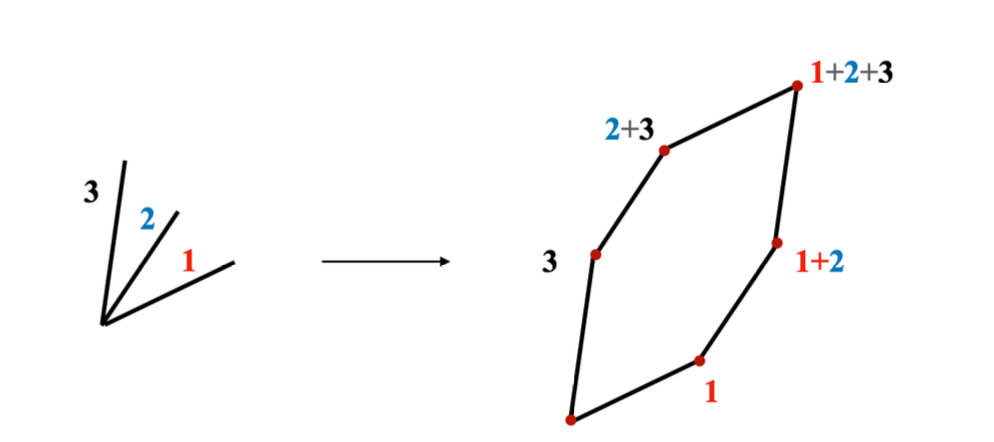}\,.$$
As one can see, the co-dimension 1 boundaries (vertices) are successively, 
\eq
{\bf v}_{1}, \;{\bf v}_{1}+{\bf v}_{2},\; {\bf v}_{1}+{\bf v}_{2}+{\bf v}_{3}, \; {\bf v}_{2}+{\bf v}_{3},\;{\bf v}_{3}\,.
\eqe
In general, the vertices are given by
\eq\label{vertices}
{\bf V}_{1,1},{\bf V}_{1,2},\ldots, {\bf V}_{1,N}, {\bf V}_{2,N},{\bf V}_{3,N},\ldots, {\bf V}_{N-1,N}\,,
\eqe
where ${\bf V}_{j_1,j_2}{=}\sum_{i=j_1}^{j_2}{\bf v}_{i}$. The co-dimension 1 boundaries defined by these vertices can be made transparent by denoting them as 
\eqa\label{nonprojpol2d}
P_j^{\textrm{lower}}&=&-\langle {\bf b}{-}{\bf V}_{1,j},{\bf v}_{j}\rangle\ge 0\,,\nonumber\\
P_j^{\textrm{upper}}&=&\langle {\bf b}{-}{\bf V}_{j,N},{\bf v}_{j}\rangle\ge 0\,.
\eqae
It is straight forward to see that due to eq.(\ref{PolytopeA}) $P_j^{\textrm{lower}, \textrm{upper}}$ is given by a sum of positive terms and therefore their vanishing is the boundary. Here upper and lower boundaries are separated by a co-dimension two boundary, a point, where all vectors are summed.

Recall that the boundary of a Minkowski sum of spaces lies within the Minkowski sum of the boundaries of the individual spaces. Here the individual space is the collection of all points on the vector, and the boundaries are the endpoints. Thus the boundaries can be identified as the Minkowski sum of the end points.

In higher dimensions the resulting $d$-dimensional polytope also has simple boundaries, generalizing eq.(\ref{nonprojpol2d}). For instance, in $d=3$, the co-dimension one boundaries, ie. $2d$ facets, can be expressed as
\eqa\label{nonprojpol3d}
P_{j_1,j_2}^{\textrm{lower}}&=&\langle {\bf b}{-}{\bf V}_{j_1,j_2},{\bf v}_{j_1},{\bf v}_{j_2}\rangle\ge 0\,,\nonumber\\
P_{j_1,j_2}^{\textrm{upper}}&=&-\langle {\bf b}{-}{\bf V}_{1,j_1}{-}{\bf V}_{j_2,N},{\bf v}_{j_1},{\bf v}_{j_2}\rangle\ge 0\,,
\eqae
for $1{\le} j_1{<}j_2{\le} N$.

\paragraph{The projective cyclic polytope}
If $0\leq \rho \leq L$, we can simply rescale all vectors ${\bf v}$ in the above by a factor of $L$. Note that in the limit $L\rightarrow \infty$ we recover the projective cyclic polytope, which is a Minkowski sum of infinite rays, instead of finite segments. To see this, the lower boundary of the 3D polytope is explicitly given by
\begin{align}
P^{\textrm{lower}}_{j_1,j_2}=\langle  {\bf b},  {\bf v}_{j_1},  {\bf v}_{j_2}\rangle - \langle {\bf V}_{j_1,j_2}, {\bf v}_{j_1}, {\bf v}_{j_2}\rangle 
\ge 0\,.
\end{align}
The first term is proportional to $L^2$, while the second to $L^3$, and therefore dominates. Since it is positive for ordered $\ell$, it automatically satisfies the inequality. However, whenever $j_2{=}j_1{+}1$ the second term vanishes, leaving only the first term, which is precisely the boundary of a cyclic polytope.

\paragraph{Equal $k$ spaces}
We are now ready to give the boundaries for the equal $k$ couplings, $(b_{k,q_1}, b_{k, q_2},\cdots )$. Here we are interested in 
\eq\label{bsu2}
b^{(s,u)}_{k,q}=\sum_{\ell} \lambda^{(s,u)}_{\ell,k,q}\int_0^1 \rho_\ell(z)z^{k+3}dz\,, \quad 0\leq \rho_\ell(z)\leq L\,.
\eqe
We can recast this as our Minkowski sum of vectors, where we have ${\mathbf b}=(b_{k,q_1}, b_{k, q_2},\cdots )$ and ${\mathbf v}_\ell=( \lambda_{\ell,k,q_1},  \lambda_{\ell,k,q_2},\cdots)$, where we've suppressed the superscript $(s,u)$. Importantly, as the $z$ integral is identical for each component of ${\mathbf b}$, it can be factored out as an effective weight for the vectors ${\mathbf v}_\ell$, 
\eq
0\leq \tilde{\rho}_{\ell}=\int_0^1 \rho_\ell(z)z^{k+3}\leq \frac{L}{k+4}\,.
\eqe 
As we will see, the $\lambda^{(s,u)}_{\ell,k,q}$ yield ordered vectors, and thus the boundaries in eq.(\ref{nonprojpol2d}) and eq.(\ref{nonprojpol3d}), after rescaling by an overall factor $\frac{L}{k+4}$, characterize the boundary for equal $k$ space.

\subsubsection{Minkowski sum of $L$-moments and unequal $k$ spaces}
As we consider a general space ($b_{k_1, q_1}$, $b_{k_2, q_2}, \cdots$), due to the unequal $k$s, the $z$ integrals in eq.(\ref{bsu2}) can no longer be considered as an effective weight as in the equal $k$ case. Instead, the space defined by: 
\eq\label{Singlemom}
b_k=\int_0^1 \rho(z) z^{k+3} dz, \quad \rho(z)\in[0,L]\,
\eqe
becomes a non-trivial region, described by the so called $L$-moment problem. We will then identify eq.(\ref{bsu2}) as the Minkowski sum of an infinite number of such $L$-moments, one for each spin, with each component weighted by $\lambda_{\ell,k,q}$. Once again, as the boundary of a Minkowski sum is given by the Minkowski sum of the boundaries, we will need to know the boundary of these individual $L$-moments.

\begin{figure}[h]
\centering
    \includegraphics[height=2in]{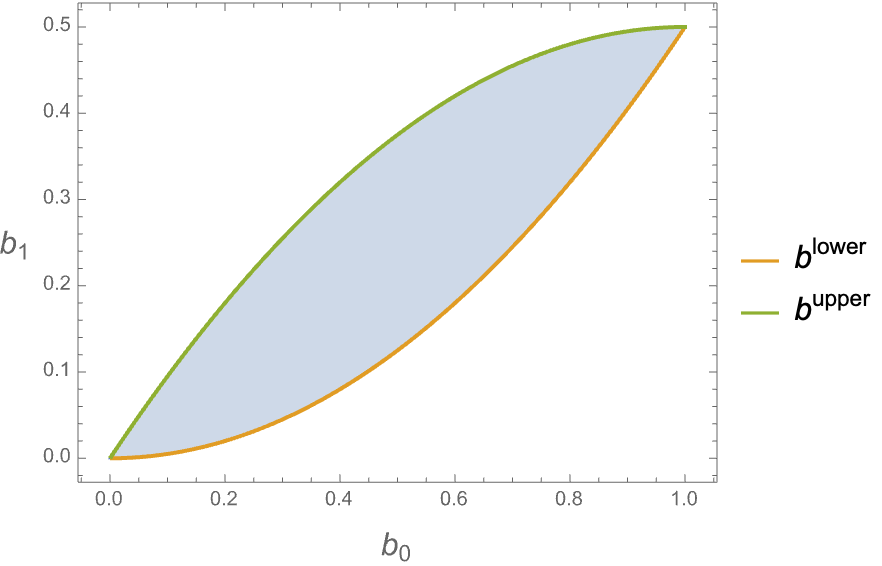} 
   \caption{$(b_0,b_1)$ space and boundaries, given in eq.(\ref{bdryeq}).}
   \label{fig:plotA1}
\end{figure}

The $L$-moment problem, i.e. delineating the necessary conditions on $b_k$s such that there exists a representation of the form in eq.(\ref{Singlemom}), is a well studied in mathematical literature (see for example \cite{PUTINAR1990288}). The boundary structure can easily be derived from our polytope discussion above. First, consider the discrete version of the integral eq.(\ref{Singlemom})
\eq
b_k=\sum_{i=1}^N \rho_i \frac{i^{k+3}}{N^{k+3}}\frac{1}{N},\quad \rho_i\in[0,L]\,.
\eqe
Then the space for ${\mathbf b}=(b_{k_1}, b_{k_2}, \cdots)$ is just a Minkowski sum of vectors, where ${\mathbf v}_i=(v_{i,k_{1}}, v_{i, k_{2}}, \cdots )$  with $v_{i,k}$=$\frac{i^{k+3}}{N^{k+3}}\frac{1}{N}$, which crucially are cyclically ordered. For two couplings, i.e. in two dimensions, we can plug these vectors into eq.(\ref{vertices}), to obtain the vertices of this space, and then take the continuous limit $N\rightarrow \infty$, to recover a simple integral\footnote{Note this is equivalent to saying the distribution $\rho(z)$ corresponding to the boundary, also known as the extremal solution, is the step function $\rho(z)=L\chi_{[0,m]}$, which is 1 for $z\in[0,m]$, and 0 otherwise.  This is similar to the extremal solution to the classical moment problem, which is given by delta functions.}

\eq
V_{1,j}^{\textrm{lower}}=L\sum_{i=1}^j \frac{i^{k+3}}{N^{k+3}}\frac{1}{N}\ \underset{N\rightarrow \infty}{\xrightarrow{\hspace*{1cm}}}\ L \int_0^{m} z^{k+3} dz=\frac{L}{k+4}m^{k+4}\,,
\eqe
where we defined $m=j/N\in[0,1]$, which becomes a continuous parameter. We find the vertices lie on two curves
\eq\label{bdryeq}
{\mathbf b}^{\textrm{lower}}=\left(\frac{L}{{k_1}{+}4} m^{k_1{+}4}, \frac{L}{{k_2}{+}4} m^{k_2{+}4}\right), \quad {\mathbf b}^{\textrm{upper}}=\left(\frac{L}{k_1{+}4} \left(1{-}m^{k_1{+}4}\right), \frac{L}{k_2{+}4} \left(1{-}m^{k_2{+}4}\right)\right)\,,
\eqe
which define the boundaries of this space. We sketch this in Figure \ref{fig:plotA1}. 

Although we will not use them in this paper, we can similarly obtain the boundaries in higher dimensions. For instance, from eq.(\ref{nonprojpol3d}) we can recover the vertices corresponding to the three-dimensional polytope, and through the same procedure obtain the 2D boundaries
\eqa
\nonumber {\mathbf b}^{\textrm{lower}}&&=\left(\frac{L}{k_1{+}4} \left(m^{k_1{+}4}{-}n^{k_1{+}4}\right), \frac{L}{k_2{+}4} \left(m^{k_2{+}4}{-}n^{k_2{+}4}\right), \frac{L}{k_3{+}4} \left(m^{k_3{+}4}{-}n^{k_3{+}4}\right)\right)\,, \nonumber\\
{\mathbf b}^\textrm{upper}&&=\left(\frac{L}{k_1{+}4} \left(1{-}m^{k_1{+}4}{-}n_1^{k{+}4}\right), \frac{L}{k_2{+}4} \left(1{-}m^{k_2{+}4}{-}n^{k_2{+}4}\right), \frac{L}{k_3{+}4} \left(1{-}m^{k_3{+}4}{-}n^{k_3{+}4}\right)\right)\,,\nonumber \\
\eqae
spanned by the parameters $0\le n\le  m\le 1$.

\begin{figure}[H] 
   \centering
        \includegraphics[height=1.8in]{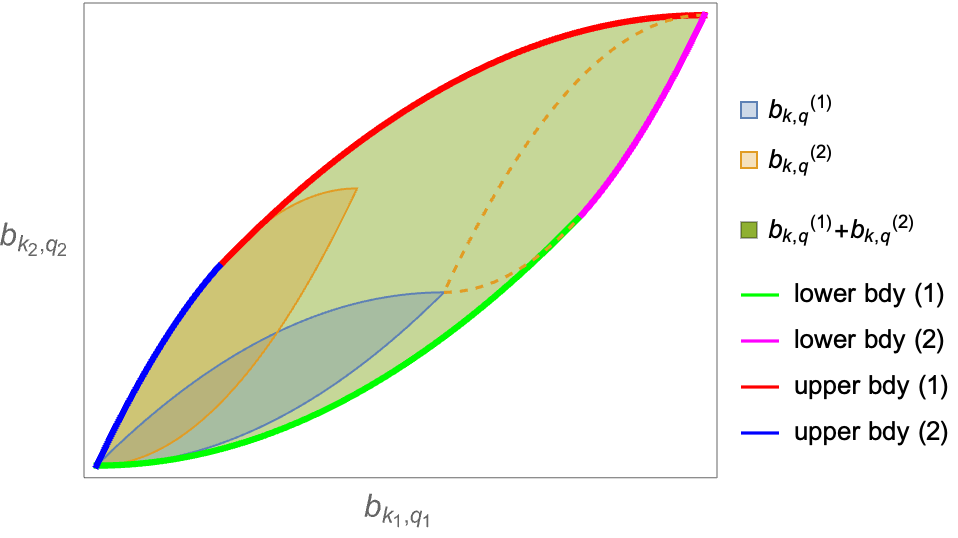} 
   \caption{The Minkowski sum of two single $L$-moments. The dashed lines represent a translation of $b_{k,q}^{(\ell_2)}$ to the point corresponding to $m_1=1$}
   \label{fig:plotM0}
\end{figure}

We now finally turn to our physical setup, which involves the following generalization of the single $L$-moment problem 
\eq
b_{k,q}=\sum_{\{\ell\}} \lambda_{\ell,k,q} \int_0^1 \rho_\ell(z)z^k dz,\quad 0\le \rho(x)\le L\,,
\eqe
where $\{\ell\}$ is the set of spins that appears in each channel. We can view the above space as a  Minkowski sum of individual single $L$-moments, rescaled by the $\lambda_{\ell,k,q}$. This sum of two such rescaled moments is sketched in Figure \ref{fig:plotM0}. We now describe the procedure in detail.

The boundaries of this Minkowski sum are completely determined by the boundaries of each rescaled single $L$-moment. Thus the problem boils down to finding the correct parameterization that gives the boundary of this sum of boundaries. For general non-equal $k$ spaces we will restrict to 2D spaces in this paper, which are much simpler to solve. This is sufficient to obtain optimal bounds on single couplings when considering one null constraint. The bounds often converge quickly as more null constraints are added, so this 2D construction is enough to extract potentially weaker bounds but analytic and of the right order of magnitude.

We begin with the 2D space $(a_{k_1,q_1},a_{k_2,q_2})$, and the Minkowski sum of two convex hulls such as those described above. Let us first assume $\lambda_{\ell,k,q}>0$. Therefore we want to find the boundaries of $a_{k,q}$ given by the Minkowski sum of two rescaled single $L$-moments:
\eq\label{singlesol}
b_{k,q}=b_{k,q}^{(\ell_1)}+b_{k,q}^{(\ell_2)}=\lambda_{\ell_1,k,q} \int_0^1 \rho_{\ell_1}(z)z^{k+3} dz+\lambda_{\ell_2,k,q} \int_0^1 \rho_{\ell_2}(z)z^{k+3} dz\,.
\eqe
As emphasized before, the boundary of the Minkowski sum is contained in the Minkowski sum of the individual boundaries. In particular, in this case we can treat the lower and upper boundaries individually.\footnote{This follows simply from the fact that the normals corresponding to each boundary live in disjoint ranges. The lower boundary has normals in $[-3/2\pi,0]$, while the upper in $[\pi/2,\pi]$.} Each individual lower boundary is a curve parameterized as
\eq
\textrm{lower bdy }[b_{k,q}^{(\ell_i)}]=\lambda_{\ell_i,k,q} \frac{L}{k+4}m_i^{k+4}\,.
\eqe
The Minkowski sum of the two boundaries will be simply 
\eq
b_{k,q}(m_1,m_2)=\frac{L}{k+4}\left(\lambda_{\ell_1,k,q} m_1^{k+4}+\lambda_{\ell_2,k,q} m_2^{k+4}\right)\,,
\eqe
which is a 2D region spanned by $m_1,m_2\in[0,1]$. The boundary of this region will then be a boundary of the Minkowski sum, and can be obtained  by minimizing $a_{k_2,q_2}$ for every $a_{k_1,q_1}$, which will give a relation between $m_1$ and $m_2$. Note that since both parameters are bounded, some boundaries can also be obtained by fixing one the parameters to either 0 or 1. The extremization can be naturally solved using a Lagrange multiplier, which in 2D boils down to solving:
\eq
\frac{ \partial b_{k_2,q_2}}{\partial m_1} \frac{\partial b_{k_1,q_1}}{\partial m_2}=\frac{ \partial b_{k_2,q_2}}{\partial m_2}\frac{\partial b_{k_1,q_1}}{\partial m_1}\,,
\eqe 
leading to the simple relation
\eq\label{rfact}
\frac{m_1}{m_2}\equiv r_{1,2}=\left(\frac{\lambda_{\ell_1,k_1,q_1}}{\lambda_{\ell_1,k_2,q_2}}\frac{\lambda_{\ell_2,k_2,q_2}}{\lambda_{\ell_2,k_1,q_1}}\right)^{\frac{1}{k_2-k_1}}\,.
\eqe
This relation implies we can express the second parameter in terms of the first, so we have one parameter for a 1D boundary as expected. The remaining potential (co-dimension 1) boundaries are obtained when one of the parameters $m_1$ or $m_2$ is equal to either $0$ or $1$. It can be easily determined (for example by requiring continuity of the boundary) the actual one is given by setting $m_1=1$ (if we assume the $\lambda_{\ell,k,q}$ are ordered, which is equivalent to having $r_{2,1}<1$). We therefore obtain that the lower boundary is given by two sections, which we can parameterize as
\eq
\ \textrm{lower bdy }[b_{k,q}]=\left\{
    \begin{array}{ll}1:\quad 
      \frac{L}{k+4}m^{k+4}\left(\lambda_{\ell_1,k,q}+r_{2,1}^{k+4}\lambda_{\ell_2,k,q}\right), & m\in [0,1] \\
   2:\quad    \frac{L}{k+4}\left(\lambda_{\ell_1,k,q}+m^{k+4} r_{2,1}^{k+4}\lambda_{\ell_2,k,q}\right), & m\in [1,r_{1,2}] 
    \end{array}
\right.
\eqe
The upper boundary can be obtained using identical reasoning. The complete Minkowski sum of two moments is illustrated in Figure \ref{fig:plotM0}. 

Let us now assume we have $N$ spins $\ell_i$, ordered such that $r_{i,1}\ge r_{i+1,1}$. Using the same steps as before we find the lower boundary is given by $N$ different sections, with section $(j)$ given by
\eqa
\nonumber b_{k,q}^{(j)}&=&  \frac{L}{k+4}\sum_{i=1}^{j-1} \lambda_{\ell_i,k,q}  +  \frac{L}{k+4}\sum_{i=j}^N \lambda_{\ell_i,k,q} \left(r_{i,1}m\right)^{k+4}\\
 &=&G^{(j)}_{k,q}+m^{k+4}F_{k,q}^{(j)}\,,
\eqae
valid for $m\in[{r_{1,j-1}},r_{1,j}]$, where $G_{k,q}^{(j)}$ and $F_{k,q}^{(j)}$
are numbers that depend only on the spin content and the specific space $(b_{k_1,q_1},b_{k_2,q_2})$ being considered. Note that through $r_{i,j}$, given in eq.(\ref{rfact}), $F_{k,q}^{(j)}$ depends explicitly on $k_1,k_2,q_1,q_2$.

\begin{figure}[H] 
   \centering
        \includegraphics[height=1.7in]{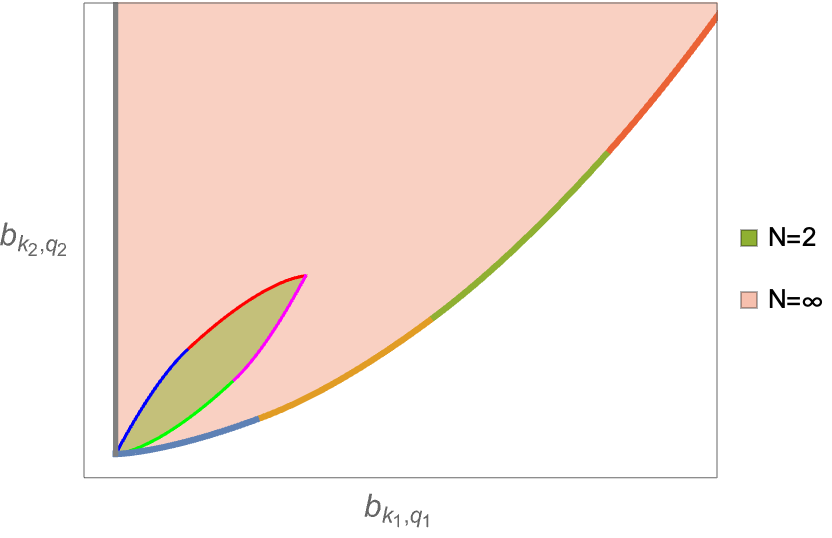} 
   \caption{Two-dimensional spaces for $N{=}2$ and $N{=}\infty$.}
   \label{fig:N}
\end{figure}

One then needs to worry if the sum in $F^{(j)}_{k,q}$ converges. If the sum diverges, this implies there are no non-projective boundaries for this space, and one only obtains projective boundaries, such  as $b_{k,q}{\ge} 0$. Convergence of this sum is therefore necessary to obtain non-projective bounds.
We sketch an example where the lower boundary converges in Figure \ref{fig:N}, where one can observe the different sections of the lower boundary.

Finally, one can find the similar upper boundary for this space. However, as we are interested in infinite number of spins, typically this boundary diverges, and so does not produce any bounds. The lower boundary will be sufficient for all the results in this paper.
\paragraph{When $\lambda_{\ell,k,q}{\le}0$}
The derivation is identical for non-positive $\lambda$ and can be easily automated, but is cumbersome to write down for completely general $\lambda$. One simply treats boundaries with slopes belonging to different quadrants independently as above, and then the complete boundary can be obtained by arranging the resulting boundaries in the correct order. This is similar to how we treated the lower and upper boundaries separately.

\subsection{A linear programming approach}
\label{LP approach}
Besides carving out the space using geometry constraints, a complementary approach to the problem is to view it as a linear functional programming problem on the spectral function $\rho_\ell(z)$. An approximate solution can be obtained by discretizing the integral along the continuous mass variable \(z\). This approach, besides being easily implemented, is fairly efficient, and allows us to both compare our analytic solution to the optimal solution, and more importantly to explore much more general spaces of couplings, including higher order null constraints. 

Suppose we want to bound the coefficient \(b_{0,0}\). Defined by eq.(\ref{bkqdef}), it is a linear functional of the spectral functions \(\rho^{++}_\ell(z)\) and \(\rho^{+-}_\ell(z)\). The boundedness of the spectral function, \(\rho_\ell(x) \leq L\), as well the null constraint, \(b_{1,1} = 0\), are again linear constraints on \(\rho_\ell(x)\). Extremizing \(b_{0,0}\) then becomes a linear functional optimization problem, which is inherently infinite-dimensional. An approximate solution can be obtained by replacing the continuous integral by a discrete Riemann sum:
\begin{equation}
    \int_0^1 \rho_\ell(z) z^k dz \rightarrow \sum_{n=1}^{N} \frac{\rho_{\ell,n}}{N} \left( \frac{n}{N} \right)^k\,,
\end{equation}
so that we have the following linear optimization problem:
\begin{align}\label{LP setup}
    \text{Maximize } \quad & b_{0,0} = \sum_{\ell=\text{even}, n}^{\ell_{max}, N} \frac{\lambda_{\ell,0,0}^{(s)} n^3}{N^4} \rho^{++}_{\ell, n} + \sum_{\ell \geq 4, n}^{\ell_{max}, N} \frac{\lambda_{\ell,0,0}^{(u)} n^3}{N^4} \rho^{+-}_{\ell, n}\,, \nonumber \\
    \text{ subject to } \quad & b_{1,1} = \sum_{\ell=\text{even}, n}^{\ell_{max}, N} \frac{\lambda_{\ell,1,1}^{(s)} n^4}{N^5} \rho^{++}_{\ell, n} + \sum_{\ell \geq 4, n}^{\ell_{max}, N} \frac{\lambda_{\ell,1,1}^{(u)}  n^4}{N^5} \rho^{+-}_{\ell, n} = 0\,, \nonumber \\
    & 0 \leq \rho^{++}_{\ell,n} \leq L^{(s)}, \text{ and } 0 \leq \rho^{+-}_{\ell,n} \leq L^{(u)}\,,
\end{align}
where \(\ell_{max}\) is the spin truncation and \(N\) the number of grids points. Following similar procedure illustrated above, one can bound arbitrary \(b_{k,q}\), or impose additional linear constraints on the spectral function, e.g., null constraints or LSD.

\subsection{Physical bounds}\label{physical}
We will now apply the previous results to obtain bounds on the couplings $b_{k,q}$. We will consider the null constraints at $k=1$ and $k=2$ separately, as the former is not valid in the presence of an $R^3$ operator. Using the results in the previous section for 2D boundaries, we will first carve out the spaces for each channel, $(b_{0,0}^{(s)},b_{1,1}^{(s)})$ and $(b_{0,0}^{(u)},b_{1,1}^{(u)})$. Next we obtain the space $(b_{0,0},b_{1,1})$ by performing the Minkowski sum of the two geometries, $(b_{0,0}^{(s)}+b_{0,0}^{(u)},b_{1,1}^{(s)}+b_{1,1}^{(u)})$. Setting the null constraint $b_{1,1}=0$ we obtain the optimal upper bound on $b_{0,0}$. Similarly we can plot the space $(b_{1,0},b_{1,1})$, which will give bounds on $b_{1,0}$. 

\begin{figure}[H] 
   \centering
    \includegraphics[height=1.8in]{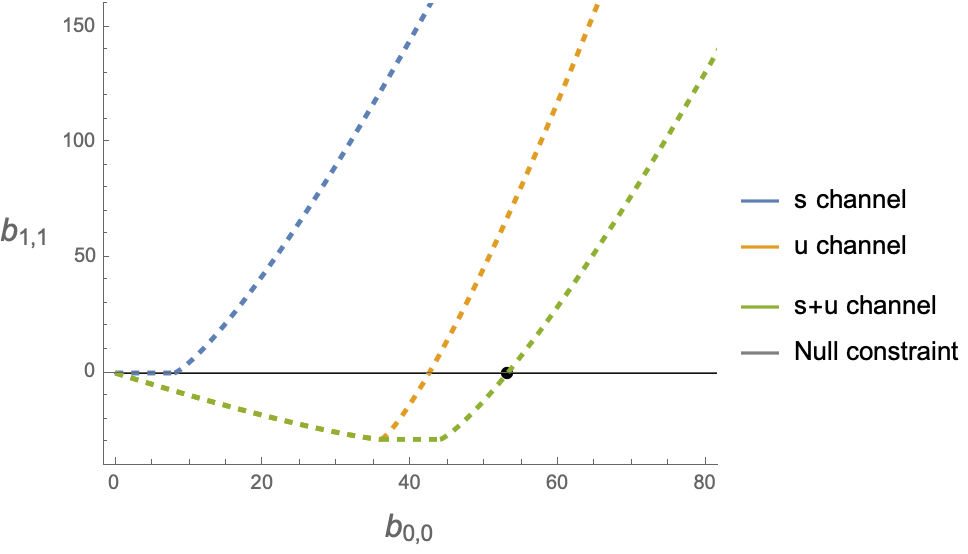} 
   \caption{Lower boundaries for  $(b^{(s,u)}_{0,0},b^{(s,u)}_{1,1})$ spaces, and their Minkowski sum, intersected with the null constraint $b_{1,1}=0$.}
   \label{fig:plotA4s}
\end{figure}

\paragraph{Bounds on $b_{0,0}$ and $b_{1,0}$ without $R^3$}
\paragraph{The  $s$-channel geometry}
For the $s$-channel we found in eq.(\ref{bkqdef}) the geometry is given by couplings
\eqa
b^{(s)}_{k,q}&=&\sum_{\ell{=}even} \lambda^{(s)}_{\ell,k,q}\int_0^1 \rho^{++}_\ell(z) z^{k+3} dz, \quad 0\le \rho^{++}_\ell(z)\le 32\,,
\eqae 
where
\eq
\lambda_{\ell,0,0}^{(s)}=(2\ell+1), \quad \lambda_{\ell,1,0}^{(s)}=(2\ell+1), \quad  \lambda_{\ell,1,1}^{(s)}=(2\ell+1)(\ell^2+\ell)\,,
\eqe
First we observe that for both spaces $(b_{0,0}^{(s)},b_{1,1}^{(s)})$ and $(b_{1,0}^{(s)},b_{1,1}^{(s)})$, $\frac{\lambda_{\ell_i,q_2}}{\lambda_{\ell_i,q_1}}\ge \frac{\lambda_{\ell_{i+1},q_2}}{\lambda_{\ell_{i+1},q_1}}$, so the spins are ordered naturally according to spin. From our previous discussion the first three sections will be given by
\eqa
\nonumber \textrm{Section 1:}&& b_{k,q}^{(s)}=\frac{L}{k+4}\lambda_{\ell_1,k,q}m^{k+4}
\,,\\
\nonumber \textrm{Section 2:} &&b_{k,q}^{(s)}=\frac{L}{k+4}\lambda_{\ell_1,k,q}+m^{k+4}F_{k,q}^{(2)}\,,\\
\textrm{Section 3:} &&b_{k,q}^{(s)}=\frac{L}{k+4}(\lambda_{\ell_1,k,q}+\lambda_{\ell_2,k,q})+m^{k+4}F_{k,q}^{(3)}\,,
\eqae
where 
\eq
F_{k,q}^{(j)}=\frac{L}{k+4}\sum_{i=j}^\infty \lambda_{\ell_i,k,q} \left(\frac{\lambda_{\ell_i,k_1,q_1}}{\lambda_{\ell_i,k_2,q_2}}\frac{\lambda_{\ell_2,k_2,q_2}}{\lambda_{\ell_2,k_1,q_1}}\right)^{k+4}\,.
\eqe
Note that the first section only contains a scalar contribution. This follows simply since $r_{1,i}=0$. We find the sums converge for this space, with the relevant results being
\eqa
\nonumber F_{0,0}^{(2)}&=&40.6353\,,\quad F_{1,1}^{(2)}=195.049\,,\\
F_{0,0}^{(3)}&=&0.635275\,,\quad F_{1,1}^{(3)}=3.04932\,.
\eqae
The three boundaries are shown in blue in Figure \ref{fig:plotA4s}. 

Next, the first three sections of the lower boundary for $(b^{(s)}_{1,0},b^{(s)}_{1,1})$ space are given by eq.(\ref{nonprojpol3d}):
\begin{align}
P_1&=b_{1,1}^{(s)}\ge 0\nonumber\,,\\
P_2&=b^{(s)}_{1,1}-6b^{(s)}_{1,0}+38.4\ge 0\nonumber\,,\\
P_3&=b^{(s)}_{1,1}-20b^{(s)}_{1,0}+576\ge 0 \,.
\end{align}
Finally, for $(b_{0,0}^{(s)},b_{1,0}^{(s)})$ space we only find the projective bounds $b_{0,0}^{(s)},b_{1,0}^{(s)},b_{0,0}^{(s)}-b_{1,0}^{(s)}\ge 0$.

\paragraph{The  $u$-channel geometry}
For the $u$-channel we have
\eqa
b^{(u)}_{k,q}&=&\sum_{\ell\ge 4} \lambda^{(u)}_{\ell,k,q}\int_0^1 \rho^{+-}_\ell(z) z^{k+3} dz, \quad 0\le \rho^{+-}_\ell(z)\le 16\,,
\eqae 
where
\eq
\lambda_{\ell,0,0}^{(u)}=(2\ell+1), \quad \lambda_{\ell,1,0}^{(u)}=-(2\ell+1), \quad  \lambda_{\ell,1,1}^{(u)}=(2\ell+1)(\ell^2+\ell-21)\,,
\eqe
We find the spins are still automatically ordered according to our criteria, but for the space $(b_{0,0}^{(u)},b_{1,1}^{(u)})$ we now encounter negative $\lambda_{\ell,1,1}$ for $\ell_1=4$. According to our previous discussion, this means the first section will only contain a contribution from this spin, while the remaining boundaries can be obtained as before, placing them where the first section ends. Explicitly, we find 
\eqa
\nonumber \textrm{Section 1:} && b_{k,q}^{(u)}=\frac{L}{k+3}(1-m^{k+4})\lambda_{\ell_1,k,q}\,,\\
\nonumber \textrm{Section 2:} &&b_{k,q}^{(u)}=\frac{L}{k+4}\lambda_{\ell_1,k,q}+m^{k+4}F_{k,q}^{(2)}\,,\\
\textrm{Section 3:} &&b_{k,q}^{(u)}=\frac{L}{k+4}(\lambda_{\ell_1,k,q}+\lambda_{\ell_2,k,q})+m^{k+4}F_{k,q}^{(3)}\,.
\eqae
where 
\eqa
\nonumber F_{0,0}^{(2)}&=&40.6353\,,\quad F_{1,1}^{(2)}=195.049\,,\\
F_{0,0}^{(3)}&=&0.635275\,,\quad F_{1,1}^{(3)}=3.04932\,.
\eqae
The three boundaries are illustrated in orange in Figure \ref{fig:plotA4s}.

For the space $(b_{1,0},b_{1,1})$, we find the following polytope conditions:
\begin{align}
P_1&=b_{1,1}^{(u)}- b_{1,0}^{(u)}\ge 0\nonumber\,,\\
P_2&=b^{(u)}_{1,1}+9b^{(u)}_{1,0}+288\ge 0\nonumber\,,\\
P_3&=b^{(u)}_{1,1}+21b^{(u)}_{1,0}+1056\ge 0 \,.
\end{align}
Finally, for $(b_{0,0}^{(u)},b_{1,0}^{(u)})$ space we also have the projective bounds $b_{0,0}^{(u)}, -b_{1,0}^{(u)},b_{0,0}^{(u)}+b_{1,0}^{(u)}\ge 0$.

\paragraph{The $(s{+}u)$-channel geometry}
Having obtained the parametric equations for the $s$ and $u$ channels, we can now compute their sum, $b_{k,q}=b_{k,q}^{(s)}+b_{k,q}^{(u)}$. Geometrically this simply means we need the Minkowski sum of the first two boundaries in Figure \ref{fig:plotA4s}, which can be computed using the same approach, leading to the green boundary in \ref{fig:plotA4s}. Intersecting with the null constraint $b_{1,1}=0$ we obtain the optimal bound on $b_{0,0}$.
Restoring the UV cutoff as the mass gap, we find 
\eqa\label{b00max}
0\le\frac{b_{0,0}}{(4\pi)^2}\le \frac{0.337}{M^8}  \,.
\eqae
Including the $(4\pi)^2$ loop factor, we observe the coupling is order unity, as expected. 

By using the polytope conditions, we can also bound $b_{1,0}$. Performing the Minkowski sum, and imposing the null condition, we find
\eq\label{b10max}
-\frac{0.202}{M^{10}} \le \frac{b_{1,0}}{(4\pi)^2} \le\frac{0.0405}{M^{10}}\,.
\eqe
We have checked that these bounds exactly match results with the linear programming method described in Section \ref{LP approach}.

\paragraph{Bounds on $b_{0,0}$ with $R^3$}
In the presence of an $R^3$ operator, the first null constraint that can be used is at $k=2$, and is given by $b_{2,1}=b_{2,2}$. To obtain optimal bounds on $b_{0,0}$, we must therefore find the boundaries for the space $(b_{0,0},b_{2,2}-b_{2,1})$. To obtain this space we treat the combination $b_{2,2}-b_{2,1}$ as a moment, which we again separate into $s$ and $u$ channel geometries, given by
\eqa\label{ncoord}
 b_{2,2}^{(s,u)}-b_{2,1}^{(s,u)}=\sum_\ell \Lambda_{\ell,2,1}^{(s,u)} \int_0^1 \rho^{(s,u)}_\ell(z)z^{5} dz\,,
\eqae
where we defined $\Lambda_{\ell,2,1}^{(s,u)}\equiv \lambda_{\ell,2,2}^{(s,u)}-\lambda_{\ell,2,1}^{(s,u)}$.
Using the same steps as before, we find a bound
\eq
\frac{b_{0,0}}{(4\pi)^2}\le \frac{0.843}{M^8}\,,
\eqe
which as expected is weaker when compared to using the $k=1$ null constraint when $R^3$ is absent in eq.\ref{b00max}, but is still order unity. 

\paragraph{Bounds on $b_{0,0}$ with maximal supersymmetry}
If we assume maximal supersymmetry, then the four graviton amplitude is part of a superamplitude, with an overall grassmann odd super-momentum delta function, i.e. it is proportional to half of the supercharge that is multiplicative in the on-shell representation:
\eq
\mathcal{M}(s,t)=\delta^{16}(Q)M(s,t)\,.
\eqe
Since for maximal SUSY, the on-shell degrees of freedom are part of a single multiplet, the function $M(s,t)$ is permutation invariant. Furthermore, as discussed in \cite{Arkani-Hamed:2022gsa}, one can show that unitarity simply translates to the residue of $M(s,t)$ is positively expanded on $D$-dimensional Gegenbauer polynomial. The dispersion relation for $b_{k,q}$, valid for $k\ge0$,  then reads \cite{Correia:2020xtr}
\eq
b_{k,q}=\frac{1}{\pi M^{d+4-2k}}\sum_\ell n_\ell^{(d)}\lambda_{\ell,k,q}\int_0^1 \rho_\ell(z) z^{\frac{d}{2}+k+1}dz\,,
\eqe
where
\eq
n_{\ell}^{(d)}=\frac{(4 \pi)^{\frac{d}{2}}(d+2 \ell-3) \Gamma(d+\ell-3)}{\pi \Gamma\left(\frac{d-2}{2}\right) \Gamma(\ell+1)}\,,\\
\eqe
and $\lambda_{\ell,k,q}$ can be obtained from the expansion
\eq
\sum_{k,q}s^{k-q} t^q\lambda_{\ell,k,q}=\mathcal{P}_{d,\ell}\left(1+2t\right)\left(\frac{1}{1-{s}}+\frac{1}{1+{s+t}}\right)\,,
\eqe
where $\mathcal{P}_{d,\ell}$ is the $d$-dimensional Gegenbauer polynomial, and can be written as:
\eq
\mathcal{P}_{d,\ell}(x)={ }_{2} F_{1}\left(-\ell, \ell+d-3, \frac{d-2}{2}, \frac{1-x}{2}\right)\,.
\eqe
Using the first null constraint $b_{2,1}=b_{2,2}$, we find the bound 
\eq
0\le\frac{b_{0,0}}{(4\pi)^5}\le \frac{91.276}{M^{14}}\,.
\eqe
Setting $M_{\rm pl}$ this applies to non-perturbative completion. It is easy to see that the IIB amplitude, $\frac{2\zeta(3)E_{{\scriptstyle \frac{3}{2}}}(\tau)}{\text{Im}[\tau]^{3/2}}\frac{1}{64\pi^7}$ lies within this region for general $\tau$.

\begin{figure}[H]
\centering
    \includegraphics[width=5in]{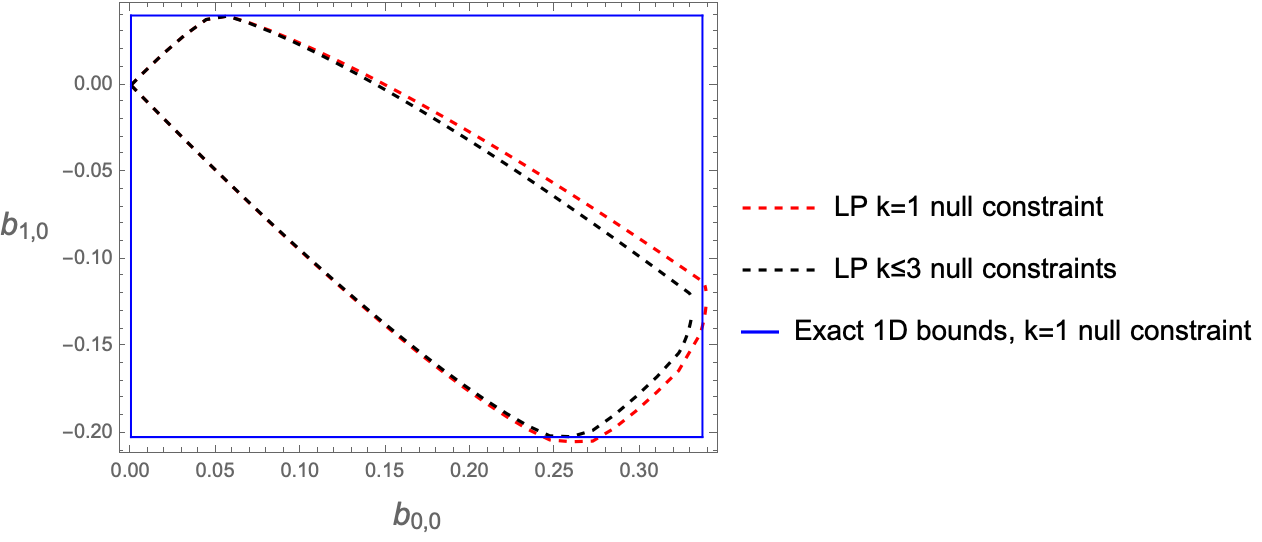}
    \caption{Non-projective space of $(b_{0,0},b_{1,0})$, comparing linear programming with exact bounds in eqn. (\ref{b00max}) and (\ref{b10max}). Including higher order null constraints only slightly reduces the space.}
    \label{b00b10space}
\end{figure}

\paragraph{Higher dimensional spaces and linear programming}
The geometrical approach described in the previous section is sufficient to describe bounds on individual couplings, and when subjected to only one null constraint. In general we would be interested in describing the space of any number of couplings, together with any (preferably very large) number of null constraints.
In principle, this can also be accomplished analytically, but would require the computation of Minkowski sums of curved objects in high dimensions, which becomes very laborious. This problem is more easily analyzed numerically, which can be done via the linear programming method described in Section \ref{LP approach}. Using this approach, we plot in Figure \ref{b00b10space} the 2D space of couplings $(b_{0,0},b_{1,0})$, when subjected up to three null constraints.\footnote{To decrease computation time for whole plot, we chose a relatively small value of $N_{max}\sim 200$, for which the bounds are weaker than optimal. This can be seen in Figure \ref{b00b10space} by comparing to the exact 1D bounds.}

\section{Low spin dominance}\label{Section4}
The solution for the $L$-moment problem can also be utilized to fully explore the consequence of low spin dominance (LSD), obtaining optimal bounds involving this condition. We begin with the dispersive representation for the couplings 
\eq
b_{k,q}=\sum_\ell \lambda_{\ell,k,q}\int_0^1 \rho_\ell(z) z^{k+3} dz ,\quad \rho_\ell(z)\ge 0\,,
\eqe
and consider the mass averaged distribution
\eq
\langle \rho_\ell \rangle_k\equiv \int_0^1 \rho_{\ell}(z) z^{k+3} dz\,.
\eqe
The LSD phenomenon is the constraint that 
\eq
\langle \rho_{\ell_1} \rangle_k\ge \alpha \langle \rho_{\ell} \rangle_k\,,
\eqe
where $\ell_1$ is the lowest spin in the spectrum, and $\alpha\ge 0$ is some parameter specifying the strength of the the LSD condition.  Since for now we only discuss equal $k$ spaces, we will drop the $k$ subscript for simplicity. Separating out the lowest spin, we can write
\eq\label{abq}
 b_{q}=\sum_\ell \langle \rho_{\ell}  \rangle \lambda_{\ell,q}= \langle \rho_{\ell_1} \rangle \lambda_{\ell,q}+\sum_{\ell> \ell_1} \langle \rho_{\ell} \rangle \lambda_{\ell,q}\,.
\eqe
We now introduce couplings $\mu_q$, 
\eq
\mu_q=\sum_{\ell>\ell_1} \langle \rho_{\ell} \rangle \lambda_{\ell,q}\,,
\eqe
which due to the LSD condition must satisfy $\langle \rho_{\ell_1} \rangle \ge \alpha \langle \rho_{\ell}\rangle $, ie. they satisfy an $L$-moment problem with $\langle\rho_\ell\rangle\le L=\frac{\langle \rho_{\ell_1}\rangle}{\alpha}$, which through eq.(\ref{abq}) will impose constraints back on $b_q$.  Since we are at equal $k$ space, the solution to this problem is given by the polytope conditions for equal $k$ spaces. Therefore to obtain bounds on the $b_q$ couplings, we merely solve for the $\mu_q$ in terms of $b_q$
\eq
\mu_q=b_q{-}\langle \rho_{\ell_1} \rangle \lambda_{\ell_1,q} \,,
\eqe
and impose the polytope constraints on the $\mu_q$
\eq\label{lsdcond}
P_{j_1,j_2,\ldots}(\mu_q)\ge 0\,,
\eqe
setting $L= \frac{\langle \rho_{\ell_1} \rangle}{\alpha}$. Unlike the previous $L$ moment problem, here the upper bound $L=\frac{\langle \rho_{\ell_1} \rangle}{\alpha}$ is not fixed beforehand. However, we can demand eq.(\ref{lsdcond}) to hold for all possible values of $\frac{\langle \rho_{\ell_1} \rangle}{\alpha}$, and this will still generate constraints.  When considering purely projective bounds, this means we simply project out $\langle \rho_{\ell_1} \rangle\ge 0$, leading to optimal constraints on $a_q$ purely in terms of $\alpha$. We can also implement the unitarity bound $\rho\le L$, which implies $\langle \rho_{\ell} \rangle_k \le \frac{L}{k+4}$ by instead projecting out $0\le \langle \rho_{\ell_1} \rangle_k\le \frac{L}{k+4} $. This will give the effect of the LSD condition on non-projective bounds. Finally, we can also combine these constraints with the full EFThedron geometry, to obtain the effect of the LSD condition on any space, including non-equal $k$.

\begin{figure}[H] 
   \centering
    \includegraphics[height=2.5in]{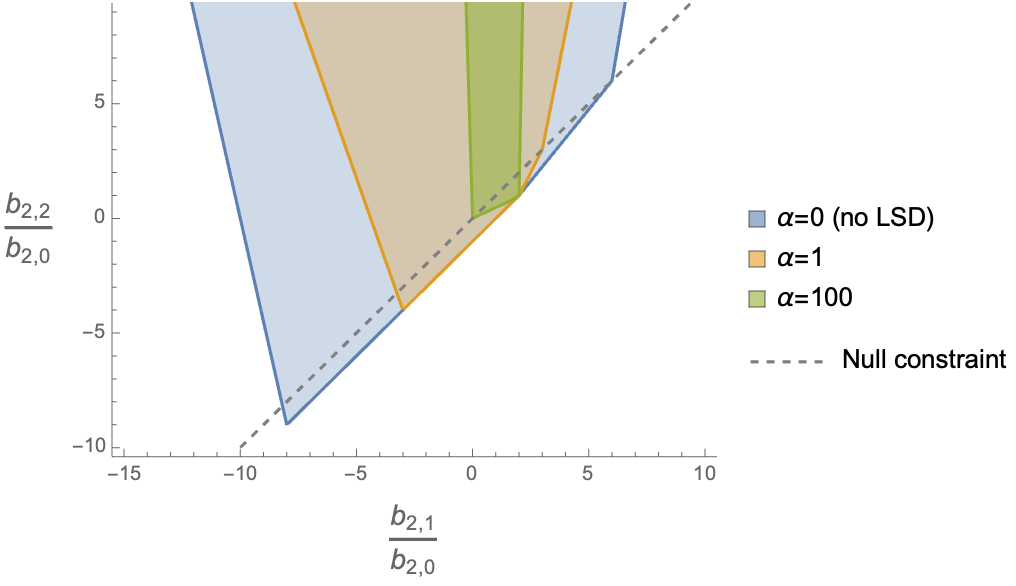} 
   \caption{The $k=2$ projective polytope for various values of $\alpha$, intersected with the null constraint $b_{2,2}=b_{2,1}$. The allowed region quickly converges beyond $\alpha=100$}
   \label{fig:plotC1}
\end{figure}

Let us first check how this approach compares with the projective bounds originally derived in Ref.~\cite{Bern:2021ppb} for the spaces $(b_{2,0},b_{2,1})$ and $(b_{4,0},b_{4,1},b_{4,2})$. For these spaces our bounds will be optimal, and we indeed find a slight improvement.
\paragraph{$k=2$ space}
For $k=2$, we use the $s-u$ channel decomposition 
\eqa
b_{2,q}=b_{2,q}^{(s)}+b_{2,q}^{(u)}\,,
\eqae
which we rewrite using the mass averaged spectral functions
\eqa
b_{2,q}^{(s,u)}=\sum_\ell \langle \rho_\ell^{(s,u)}\rangle \lambda^{(s,u)}_{\ell,q}\,.
\eqae
It can be checked that both $\lambda^{(s,u)}_{\ell,q}$ are cyclically ordered, and so we can use the polytope conditions in eq.(\ref{nonprojpol2d}) and (\ref{nonprojpol3d}),
\eqa
&&\nonumber P^{(s)}_{i}\left(b^{(s)}_{2,q}-\langle \rho_{\ell_1}^{(s)}\rangle\, \lambda^{(s)}_{\ell_1,q}\right)\ge 0\,,\\
&&\nonumber P^{(s)}_{i,j}\left(b^{(s)}_{2,q}-\langle \rho_{\ell_1}^{(s)}\rangle\, \lambda^{(s)}_{\ell_1,q}\right)\ge 0\,,
\eqae
as well as the projective constraint $b_{2,0}^{(s)}\ge 0$, and similar for the $u$ channel. We have for $s$ channel $\ell^{(s)}_1=0$, $L^{(s)}=\frac{\langle\rho_{\ell_1}^{(s)}\rangle }{\alpha}$ and  for the $u$ channel $\ell^{(u)}_1=4$, $L^{(u)}=\frac{\langle\rho_{\ell_1}^{(u)}\rangle }{\alpha}$. While the polytope conditions involve an infinite number of spins, in practice truncating to some not very large spin $\ell\sim 20$ is sufficient due to the null constraint intersecting only some facets. Using these constraints we first carve out the projective polytope for $(\frac{b_{2,1}}{b_{2,0}},\frac{b_{2,2}}{b_{2,0}})$. As expected, we find the LSD condition imposes constraints on the polytope, even in the absence of a null constraint. We plot this in Figure \ref{fig:plotC1}.
Imposing the null constraint $b_{2,2}=b_{2,1}$, we find the bounds on the ratio $b_{2,1}/b_{2,0}$, for various $\alpha$
\eqa
\nonumber \alpha=1&: &\quad-3.257\le \frac{b_{2,1}}{b_{2,0}}\le 3\,,\\
\nonumber\alpha=100&:&\quad0\le \frac{b_{2,1}}{b_{2,0}}\le 2.019\,,\\
\alpha=\infty&:& \quad0\le \frac{b_{2,1}}{b_{2,0}}\le 2\,.
\eqae
We compare these to the original findings in Ref.~\cite{Bern:2021ppb},
\eqa
\nonumber \alpha=1&:&\quad -4.153\le \frac{b_{2,1}}{b_{2,0}}\le 3\,,\\
\nonumber \alpha=100&:&\quad  -0.083\le \frac{b_{2,1}}{b_{2,0}}\le 2.286\,,\\
\alpha=\infty&:&\quad 0\le \frac{b_{2,1}}{b_{2,0}}\le 2\,,
\eqae
and observe that for finite $\alpha$ the $L$-moment optimal solution provides a slight improvement.

\begin{figure}[H] 
   \centering
    \includegraphics[height=2.7in]{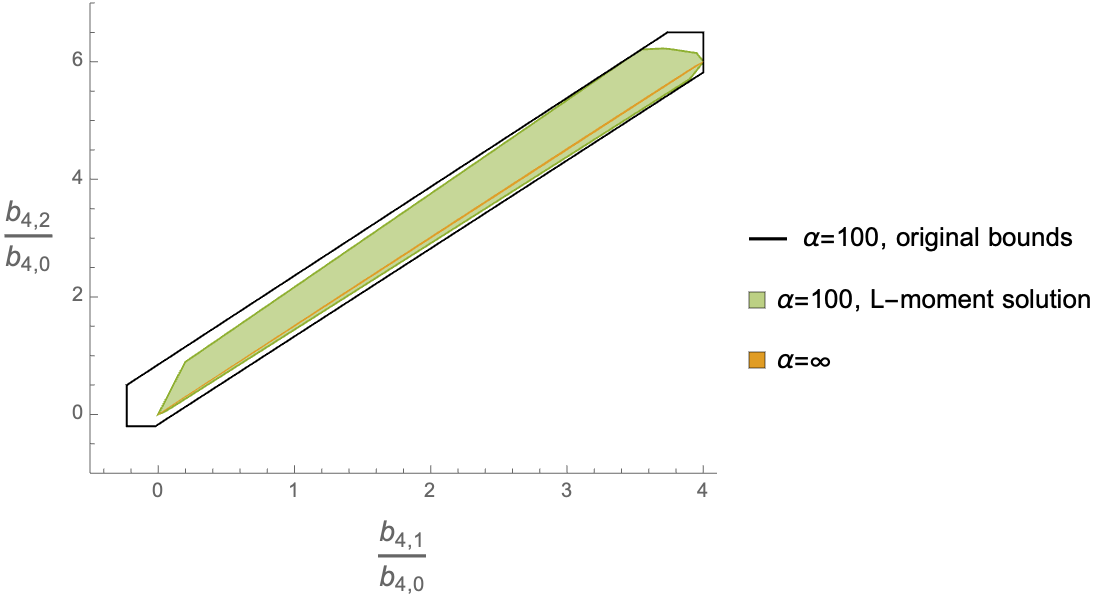} 
   \caption{The $k=4$ projective polytope after being intersected with two null constraints $b_{4,3}=2(b_{4,2}-b_{4,1})$ and $b_{4,4}=b_{4,2}-b_{4,1}$, for $\alpha=100$ and $\infty$. We observe the optimal bounds from the $L$-moment problem solution are slightly better than the previously derived bounds at $\alpha=100$, but identical for $\alpha=\infty$.}
   \label{fig:plotC2}
\end{figure}

\paragraph{$k=4$ space}
At $k=4$ the space we must carve out is $(b_{4,0},b_{4,1},b_{4,2},b_{4,3},b_{4,4})$, therefore we must use all polytopes up to the 5D one. We then intersect with the null planes $b_{4,3}=2(b_{4,2}-b_{4,1})$ and $b_{4,4}=b_{4,2}-b_{4,1}$. We obtain the projective space for  $(\frac{b_{4,1}}{b_{4,0}},\frac{b_{4,2}}{b_{4,0}})$ shown in Figure \ref{fig:plotC2}. As before the bounds for finite $\alpha$ are slightly stronger than the ones previously found in Ref.~\cite{Bern:2021ppb}.

\begin{figure}[H] 
   \centering
   \includegraphics[width=\linewidth]{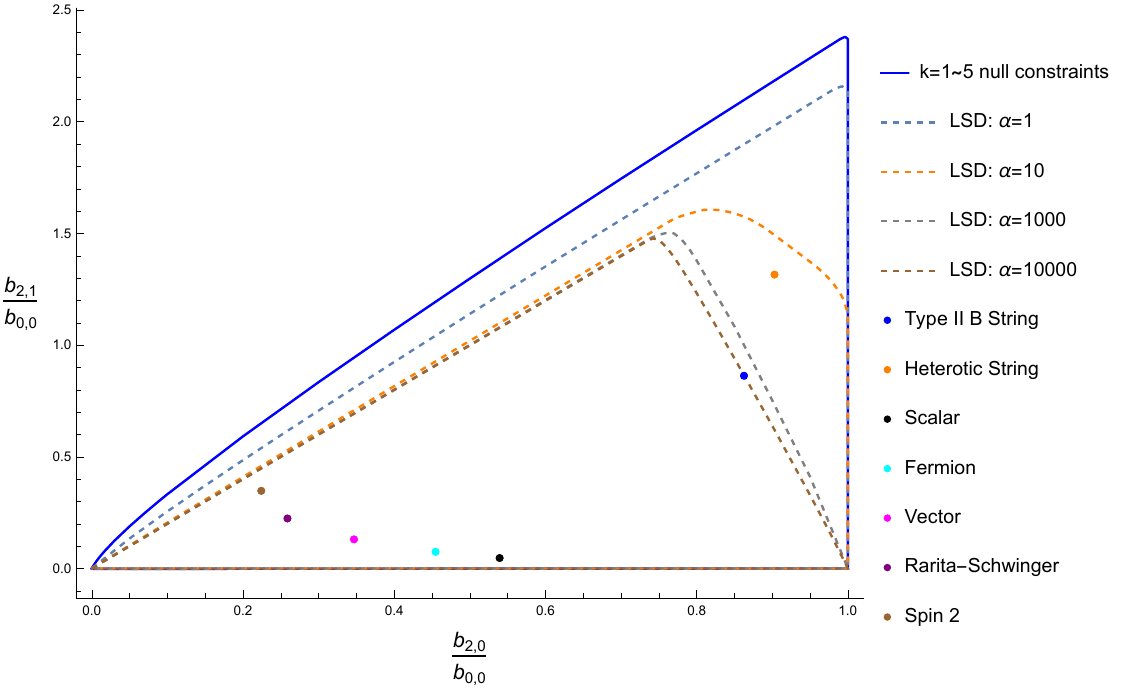} 
   \caption{The \(b_2\) projective space, normalized by \(b_{0,0}\). The boundaries corresponding to different LSD parameters are obtained using linear programming, with \(k \leq 5\) null constraints and \(k = 2\) LSD conditions imposed. }
   \label{b2-LSD-plot}
\end{figure}

\paragraph{Non equal $k$ space}
Let us consider the \(k = 2\) space, but normalized by \(b_{0,0}\), and assume the absence of \(R^3\) operator so that we have access to all the null constraints including \(k = 1\). The \(k=2\) LSD conditions are imposed and the space is carved out using linear programming. The boundaries corresponding to various different LSD parameters \(\alpha\) are shown in Figure \ref{b2-LSD-plot}. As \(\alpha\) increases, the allowed space converges to some smaller region, where the known theories are known to exist. Note that heterotic string theory in known to only satisfy the LSD condition with $\alpha=10$, which is consistent with our findings.

Finally, we point out that, as also discussed in Ref.~\cite{Bern:2021ppb}, if a theory satisfies LSD it must live in the space carved out above, but the converse is not necessarily true. 

\section{Conclusion and outlook}\label{conclusion}
In this paper we considered the space of consistent gravitational EFT by carving out the allowed region for EFT coefficients of the $D^{2n}R^4$ operator, using its dispersive representation and imposing UV unitarity and IR symmetry. Using numerical SDP we find that, while the region converges rapidly as we extend the consistency with crossing symmetry to higher order in derivative expansion, the boundary spectrum is populated by higher spins states with mass exactly at the mass-gap. This points to an artifact of the numerical approach. It would also be interesting to know what is wrong physically about such a spectrum or  how to remove such spectrums in a systematic manner. Perhaps one can eventually show that such spectrums cannot admit a consistent massive Compton amplitude, where the massive states of the accumulation point are on the outside, along the lines of \cite{Chen:2021bvg}.

Numeric SDP methods lead to projective bounds on ratios of EFT coupling, which show that in general the relative size of EFT coefficients cannot be arbitrarily large outside of naive power counting, as expected. However, one also expects that individual coefficients must be $\mathcal{O}(1)$ with respect to the UV cutoff. To demonstrate this we must derive non-projective bounds utilizing the fact that the spectral function $\rho$ is bounded from above. To this end we utilize our analytic characterization of the dispersion relation in terms of geometry, which allows us to map the upper bound of $\rho$ to the single $L$-moment problem in mathematical literature. Importing its solution and extending to a bivariate $L$ moment problem allows us to derive sharp bounds in agreement with the $\mathcal{O}(1)$ expectation. This analytic approach can then be easily extended to derive optimal bounds originating from the low spin dominance condition.

Note that constraints on $D^{2n}R^4$ using forward limit dispersion relation cannot delineate theories that UV complete Einstein-Hilbert action from those that simply arrises from integrating low scale massive states that do not participate in the UV completion of perturbative gravity. To do so one needs to consider bounding operators proportional to $s^2$, which are subject to the influence of the $t$-channel graviton pole. As mentioned in the introduction this can be circumvented by considering the finite impact parameter space dispersive representation introduced in~\cite{Caron-Huot:2021rmr}. Thus it will be important to derive non-projective bounds for such construction, and t would be interesting if such bounds also have a geometric and analytic interpretation in terms of the moment problem.

Finally, we considered $D=4$ where graviton scattering can be expanded on unique orthogonal polynomials. In $D>4$ this is no longer true, and one needs to expand on a matrix basis whose size is determined by the number of distinct three point couplings of external gravitons to irreps of SO($D{-}1$). Recently it was shown that gauge invariance leads to a simple and finite size matrix irrespective of the irrep~\cite{Arkani-Hamed:2022gsa}. This will open the way to constraining higher dimensional quantum gravity where one is free of obstructions from IR divergences.  

Our analysis is by no means exhaustive. One can next combine the various non-projective constraints from the unitarity bound, the low spin dominance condition applied across different $k$ orders, together with different helicity sectors and associated null constraints, in order to carve out a multitude of EFT spaces with different assumptions. It would be interesting to see how much the allowed space can be further reduced when all possible constraints are imposed.

Another approach to obtaining bounds on EFT couplings, using geometric function theory, has been recently developed in Refs.~\cite{Haldar:2021rri,Zahed:2021fkp,Raman:2021pkf,Chowdhury:2021ynh}. Here one trades manifest locality for manifest crossing symmetry, and uses so-called Bieberbach bounds to constrain EFT coefficients. It would be fascinating to see how this can be related to the geometry of the EFThedron.

On the mathematical side, the bi-variate $L$-moment problem for two continuous moments has been solved in various instances \cite{PUTINAR1990288} through another generalization of the moment matrix into a moment kernel. However, it is not clear if one can import those results and apply them to our continuous $\times$ discrete $L$-moment problem. The advantage would be finding inequalities directly in terms of the couplings, avoiding the need to perform the Minkowski sum of complicated boundary structures.

\acknowledgments 
 L-y Chiang, Y-t Huang, and H-c Weng are supported by Taiwan Ministry of Science and Technology Grant No. 109-2112-M-002 -020 -MY3.   L Rodina is supported by Taiwan Ministry of Science and Technology Grant No. 109-2811-M-002-523.

\appendix 
\section{Explicit details for SDP}\label{SDPApp}

In this section, we will show how to derive bounds in $\mathbb{P}^2$ space and demonstrate a simple example of bounding $\frac{b_{4,1}}{b_{4,0}}$. To derive bounds in $\mathbb{P}^2$ space, $\left(\frac{b_{k_2,q_2}}{b_{k_1,q_1}},\frac{b_{k_3,q_3}}{b_{k_1,q_1}}\right)$, one should first find $\mathbb{P}^1$ bound $A{\leq} \frac{b_{k_2,q_2}}{b_{k_1,q_1}}{\leq} B$ then find the upper and lower bounds on $\frac{b_{k_3,q_3}}{b_{k_1,q_1}}$ when fixing $\frac{b_{k_2,q_2}}{b_{k_1,q_1}}{=}r{\in}\left[A,B\right]$. This can be simply done by adding another component in \(\vec{F}_a\):
\begin{equation}
    \vec{F}_{m_a,\ell_a} = 
	\begin{pmatrix}
		\frac{B_{k_1, q_1} (\ell_a) }{ m_a^{2(k_1 + 1)} }\\
		\frac{B_{k_2, q_2} (\ell_a) }{ m_a^{2(k_2 + 1)} }\\
		\frac{B_{k_3, q_3} (\ell_a) }{ m_a^{2(k_2 + 1)} }\\
		\frac{N_k(\ell_a)}{m_a^{2(k+1)}}\\
		\vdots
	\end{pmatrix}.
\end{equation}
$B_{k_3,q_3}$ is a three-by-three matrix defined by the dispersive representation of $b_{k_3,q_3}$. The SDP optimization problem that we are solving are: 
\begin{itemize}
    \item Finding the best upper bound $\left(\frac{b_{k_3,q_3}}{b_{k_1,q_1}}\leq A',\ \text {given }\frac{b_{k_2,q_2}}{b_{k_1,q_1}}=r\right)$ is equivalent to finding a $D{+}3$ dimensional $\vec{v}$ such that,
    \eq
  (0,0,{-}1,0,\cdots)\cdot\vec{v}=1,\; \&\quad \tilde{\vec{F}}_{x,\ell}^T\cdot \vec{v}\succeq 0\,\quad \forall x \geq 0, \, \ell=0,1,...,\ell_{max}\,,
    \eqe
    and minimize $ A'= (1,r,0,\cdots) \cdot \vec{v}$.
       \item Finding the best lower bound $\left(B'\geq\frac{b_{k_3,q_3}}{b_{k_1,q_1}},\ \text {given }\frac{b_{k_2,q_2}}{b_{k_1,q_1}}=r\right)$ is equivalent to 
        \eq
  (0,0,1,0,\cdots)\cdot\vec{v}=1,\; \&\quad \tilde{\vec{F}}_{x,\ell}^T\cdot \vec{v}\succeq 0\,\quad \forall x \geq 0, \, \ell=0,1,...,\ell_{max}\,,
    \eqe
    and minimizing $B'= ({-}1,{-}r,0,\cdots) \cdot \vec{v}$.
\end{itemize}
By scanning through the region $r\in[B,A]$, one will obtain the desired $\mathbb{P}^2$ space boundary for the $\left(\frac{b_{k_2,q_2}}{b_{k_1,q_1}},\frac{b_{k_3,q_3}}{b_{k_1,q_1}}\right)$ . One can also generalize this method and carve out the boundary for $\mathbb{P}^3$, $\mathbb{P}^4...$ spaces.

We now give a simple example of bounding $\mathbb{P}^1$ space, $\frac{b_{4,1}}{b_{4,0}}$. First, from eq.(\ref{dispersion++--}), the  dispersive representation of the Wilson coefficients $b_{4,0}$ and $b_{4,1}$ are:
\begin{align}\label{b40b41b42disprep}
b_{4,0}&=\sum_{\ell_a=even}\frac{|p_a^{++}|^2}{m_a^{18}}+\sum_{\ell_b\geq 4}\frac{|p_b^{+-}|^2}{m_b^{18}}\nonumber\\
b_{4,1}&=\sum_{\ell_a=even}\frac{\ell_a(\ell_a{+}1)|p_a^{++}|^2}{m_a^{18}}+\sum_{\ell_b\geq 4}\frac{-(\ell_b^2{+}\ell_b{-}24)|p_b^{+-}|^2}{m_b^{18}}.
\end{align}
The crossing symmetry conditions eq.(\ref{tunullconstraintk=4plot}) at $k{=}4$ have dispersive representations:
\begin{align}
    &n_1=-b_{4,3} + 2 b_{4,4} = \sum_{\ell_a\geq4}\frac{(\ell_a-4) (\ell_a-2) (\ell_a-1) \ell_a (\ell_a+1) (\ell_a+2) (\ell_a+3) (\ell_a+5) |p_a^{+-}|^2}{288 m^{18}_a} \nonumber\\& +\sum_{\ell_b=even}\frac{(\ell_b (\ell_b+1) (\ell_b (\ell_b+1) ((\ell_b-12) \ell_b (\ell_b+1) (\ell_b+13)+8588)-197712)+1606464) |p_b^{++}|^2}{288 m^{18}_b}= 0 ,\nonumber\\  &n_2=b_{4,1} - b_{4,2} + b_{4,4} =\sum_{\ell_a\geq4}\frac{(\ell_a-4) (\ell_a-2) \ell_a (\ell_a+1) (\ell_a+3) (\ell_a+5) \left(\ell_a^2+\ell+6\right) |p_a^{+-}|^2}{576 m^{18}_a} \nonumber\\&+ \sum_{\ell_b=even}\frac{\left(\ell_b (\ell_b+1) \left(\ell_b (\ell_b+1) \left(\ell_b (\ell_b+1) \left(\ell_b^2+\ell_b-164\right)+9324\right)-219024\right)+1802304\right) |p_b^{++}|^2}{576 m^{18}_b}=0\,.
\end{align}
we organize the dispersive representation can be organized into the following matrix form in order to implement it into SDP problem:
\begin{align}
&\sum_{\ell_a{=}0,2}
\begin{pmatrix}
p^{++}_a & p^{--}_a & p^{+-}_a
\end{pmatrix}
\begin{pmatrix}
B_{4,0}(\ell_a)\\
B_{4,1}(\ell_a)\\
N_1(\ell_a)\\
N_2(\ell_a)
\end{pmatrix}
\begin{pmatrix}
p^{*++}_a\\
p^{*--}_a\\
p^{*+-}_a
\end{pmatrix}
+
\sum_{\ell_b\geq 4,\ell_b{=}even}
\begin{pmatrix}
p^{++}_b & p^{--}_b & p^{+-}_b
\end{pmatrix}
\begin{pmatrix}
\tilde{B}_{4,0}(\ell_b)\\
\tilde{B}_{4,1}(\ell_b)\\
\tilde{N}_1(\ell_b)\\
\tilde{N}_2(\ell_b)
\end{pmatrix}
\begin{pmatrix}
p^{*++}_b\\
p^{*--}_b\\
p^{*+-}_b
\end{pmatrix}\nonumber\\
&\sum_{\ell_c\geq 4,\ell_c{=}odd}
\begin{pmatrix}
p^{++}_c & p^{--}_c & p^{+-}_c
\end{pmatrix}
\begin{pmatrix}
B'_{4,0}(\ell_c)\\
B'_{4,1}(\ell_c)\\
N'_1(\ell_c)\\
N'_2(\ell_c)
\end{pmatrix}
\begin{pmatrix}
p^{*++}_c\\
p^{*--}_c\\
p^{*+-}_c
\end{pmatrix}
=
\begin{pmatrix}
b_{4,0}\\
b_{4,1}\\
0\\
0
\end{pmatrix}.
\end{align}
The corresponding $B$ matrices come from the Wilson coefficients $b_{4,0}$ and $b_{4,1}$,
\begin{align}
& B_{4,0}=\begin{pmatrix}1 & 0 & 0\\ 0 & 0 & 0 \\ 0 & 0 & 0\end{pmatrix},\ \tilde{B}_{4,0}=\begin{pmatrix}1 & 0 & 0\\ 0 & 0 & 0 \\ 0 & 0 & 1\end{pmatrix},\ B'_{4,0}=\begin{pmatrix}0 & 0 & 0\\ 0 & 0 & 0\\ 0 & 0 & 1\end{pmatrix},\nonumber\\ &B_{4,1}=\begin{pmatrix}\ell(\ell+1) & 0 & 0\\0 & 0 & 0\\ 0 & 0 & 0\end{pmatrix},\ \tilde{B}_{4,1}=\begin{pmatrix}\ell(\ell+1) & 0 & 0\\0 & 0 & 0\\ 0 & 0 & -(\ell^2+\ell-24)\end{pmatrix},\ B'_{4,1}\begin{pmatrix}0 & 0 & 0\\0 & 0 & 0\\ 0 & 0 & -(\ell^2+\ell-24)\end{pmatrix}\,.
\end{align}
The $N$ matrices are obtained from the null constraints and they are defined below:
\begin{align}
& N_1=\begin{pmatrix}n_1 & 0 & 0 \\ 0 & 0 & 0\\ 0 & 0 & 0\end{pmatrix},\ 
\tilde{N}_1=\begin{pmatrix}n_1 & 0 & 0 \\ 0 & 0 & 0\\ 0 & 0 & n_1'\end{pmatrix},\ 
N_1'=\begin{pmatrix}0 & 0 & 0\\ 0 & 0 & 0\\ 0 & 0 & n_1'\end{pmatrix},\nonumber\\
& N_2=\begin{pmatrix}n_2 & 0 & 0\\ 0 & 0 & 0\\0 & 0 & 0\end{pmatrix},\ 
\tilde{N}_2=\begin{pmatrix}n_2 & 0 & 0\\ 0 & 0 & 0\\0 & 0 & n_2'\end{pmatrix},\ 
N_2'=\begin{pmatrix} 0 & 0 & 0\\ 0 & 0 & 0\\ 0 & 0 & n_2'\end{pmatrix}\,,\nonumber\\
& n_1=(\ell (\ell+1) (\ell (\ell+1) ((\ell-12) \ell (\ell+1) (\ell+13)+8588)-197712)+1606464)\,,\nonumber\\
& n'_1=(\ell-4) (\ell-2) (\ell-1) \ell (\ell+1) (\ell+2) (\ell+3) (\ell+5)\,,\nonumber\\
& n_2=\left(\ell (\ell+1) \left(\ell (\ell+1) \left(\ell (\ell+1) \left(\ell^2+\ell-164\right)+9324\right)-219024\right)+1802304\right)\,,\nonumber\\
& n_2'(\ell-4) (\ell-2) \ell (\ell+1) (\ell+3) (\ell+5) \left(\ell^2+\ell+6\right)\,.
\end{align}
Note that there is a factor of $1/m^{18}$ that is absorbed into $p_a$'s. It is now suitable to define the vector
\begin{equation}
\vec{F}_\ell=\begin{pmatrix}
B_{4,0} \\ B_{4,1} \\ N_1 \\ N_2
\end{pmatrix}\ (\ell=0,2)\quad 
\vec{F}_\ell=\begin{pmatrix}
\tilde{B}_{4,0} \\ \tilde{B}_{4,1} \\ \tilde{N}_1 \\ \tilde{N}_2
\end{pmatrix}\ (\ell=4,6, 8...)\quad 
\vec{F}_\ell=\begin{pmatrix}
B'_{4,0} \\ B'_{4,1} \\ N'_1 \\ N'_2
\end{pmatrix}\ (\ell=5,7,9...),
\end{equation}
and set up the SDP problem to give a upper and lower bounds on the ratio $\frac{b_{4,1}}{b_{4,0}}$. The SDP problem for the upper bound is to find a vector $\vec{v}$ such that
\begin{equation}
(0,-1,0,0)\cdot\vec{v}=1,\& \quad \vec{F}_\ell^T\cdot\vec{v}\succeq0, \quad \ell=0,2,4,5,6...\ell_{max}.
\end{equation}
and one minimizes $A=(1,0,0,0)\cdot\vec{v}$. The SDP problem for the lower bound is to find a vector $\vec{v}$ such that
\begin{equation}
(0,1,0,0)\cdot\vec{v}=1,\& \quad \vec{F}_\ell^T\cdot\vec{v}\succeq0, \quad \ell=0,2,4,5,6...\ell_{max}.
\end{equation}
and one maximizes $A=(1,0,0,0)\cdot\vec{v}$. Using the solver SDPB, we are able to find the bound $-18.887<\frac{b_{4,1}}{b_{4,0}}<20$. One may notice that the matrices that we considered above are diagonal and it may not be necessary to formulate the problem into three-by-three matrices, however, when one considers bounding nulls constraints aside from MHV sectors, such as the projective space $\left(\frac{a_{4,0}}{b_{0,0}},\frac{a_{5,1}}{b_{0,0}}\right)$ that we have considered in section \ref{explict_projective_region}, off-diagonal terms will show up, thus, it is necessary to consider such three-by-three matrices.

\section{Convergence analysis on the linear programming approach}
\label{LP}
Introduced in Section \ref{LP approach}, the problem of bounding Wilson coefficients can be regarded as a linear functional programming problem, and approximate solutions can be obtained by truncating the spins and discretizing the integral with respect to mass. Convergence to the exact solution is expected as \(\ell_{max},N \rightarrow \infty\), while for the 2D regions considered in this paper, truncation/discretization parameters of order \( 10^2 \sim 10^3\) will typically be sufficient to give satisfactory results (\(\sim \pm 1 \%\) around the convergence value), which can be handled efficiently by an average personal computer using the \textbf{FindMaximum} function in Mathematica. Here we consider two examples to demonstrate this.

\begin{figure}[H]
\centering
    \includegraphics[width=0.8\linewidth]{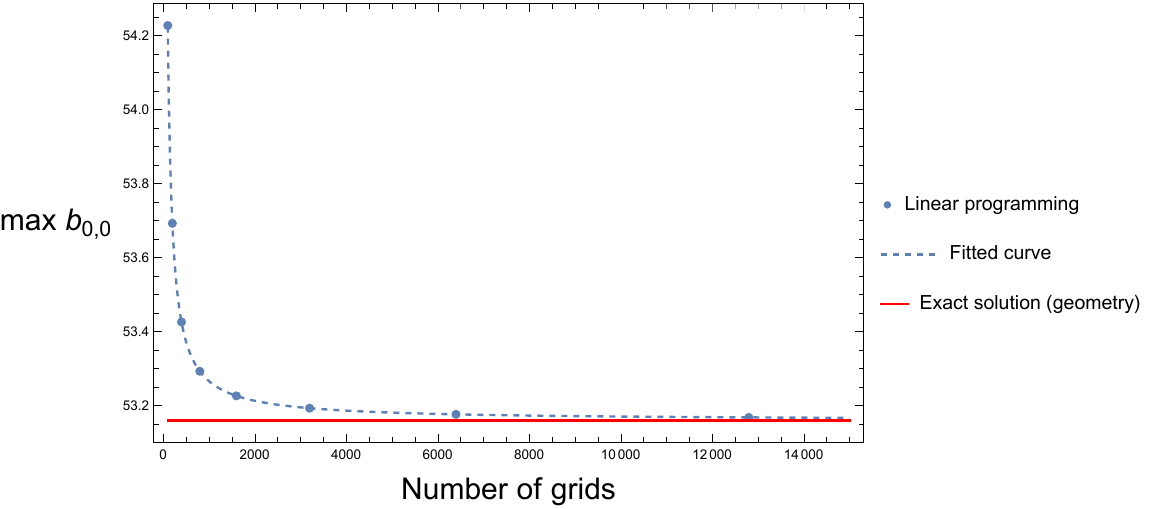}
    \caption{Convergence of \(b_{0,0}\) as the number of grids increases.}
    \label{b00_vs_N}
\end{figure}

Table \ref{b00vslN} shows the influence of \(\ell_{max}\) and \(N\) on the maximal value of \(b_{0,0}\), with \(k = 1\) null constraint imposed. The value approaches the exact solution eq.(\ref{b00max}) as the parameters increase, while for this particular case it is insensitive to the spin truncation as the spin exceeds \(20\). Indeed, with crossing symmetry imposed, spins much higher than the zeros of the null constraint will have power-law suppressed spectral function, and in eq.(\ref{LP setup}) the polynomial \(\ell^2+\ell-21\) has root \(\ell \approx 4.11\). Plotting \(\max b_{0,0}\) against \(N\), one observes a convergence toward the exact solution in Figure \ref{b00_vs_N}. For sufficiently high \(N\), the error decays roughly as \(N^{-1}\), illustrated by the fitted curve. Extrapolation then predicts an upper bound \(b_{0,0} \leq 53.1594\), slightly smaller than the analytical solution \(b_{0,0} \leq 53.1595\).

\begin{table}[H]
    \centering
    \caption{Maximal value of \(b_{0,0}\) obtained with different optimization parameters. Notice its insensitiveness to spin truncation.}
    \begin{tabular}{|l|l|l|l|l|l|}
        \hline
        \(\ell_{max}\) / \(N\) & 100     & 200     & 400     & 800     & 1600    \\ \hline
        10				    & 54.2270 & 53.6923 & 53.4255 & 53.2924 & 53.2259 \\ \hline
        20                  & 54.2271 & 53.6924 & 53.4256 & 53.2924 & 53.2260 \\ \hline
        100                 & 54.2271 & 53.6924 & 53.4256 & 53.2924 & 53.2260 \\ \hline
    \end{tabular}
    \label{b00vslN}
\end{table}

\bibliography{mybib}{}
\bibliographystyle{JHEP}
\end{document}